\documentclass[aps]{revtex4}%
\usepackage{amsfonts}
\usepackage{amsmath}
\usepackage{amssymb}
\usepackage{epsfig}
\usepackage{graphicx}
\usepackage{hyperref}
\begin{document}
\title{Fluid-crystal coexistence for proteins and inorganic nanocolloids:
dependence on ionic strength}
\author{Peter Prinsen and Theo Odijk*,}
\affiliation{Complex Fluids Theory, Faculty of Applied Sciences, Delft University of
Technology, Delft, the Netherlands}

\begin{abstract}

We investigate theoretically the fluid-crystal coexistence of
solutions of globular charged nanoparticles like proteins and
inorganic colloids. The thermodynamic properties of the fluid
phase are computed via the optimized Baxter model. This is done
specifically for lysozyme and silicotungstates for which the bare
adhesion parameters are evaluated via the experimental second
virial coefficients. The electrostatic free energy of the crystal
is approximated by supposing the cavities in the interstitial
phase between the particles are spherical in form. In the
salt-free case a Poisson-Boltzmann equation is solved to calculate
the effective charge on a particle and a Donnan approximation is
used to derive the chemical potential and osmotic pressure in the
presence of salt. The coexistence data of lysozyme and
silicotungstates are analyzed within this scheme, especially with
regard to the ionic-strength dependence of the chemical
potentials. The latter agree within the two phases provided some
upward adjustment of the effective charge is allowed for.

\vspace{10pt}

*Address for correspondence: T. Odijk, P.O. Box 11036, 2301 EA
Leiden, the Netherlands

E-mail: odijktcf@wanadoo.nl

\end{abstract}
\maketitle

\section*{I. INTRODUCTION}

One current view of protein crystallization centers on the second
virial coefficient $B_{2}$ being a relevant quantity determining
the onset of crystallization \cite{GE1,GE2,GUO,HA1}. There exists
a crystallization slot of negative $B_{2}$ values which expresses
a necessary range of solution conditions for adequate crystals to
grow. A negative value of $B_{2}$ implies a Baxter stickiness
parameter as it is conventionally defined and we here denote by
$\tau_{0}$. Thus, in a similar vein, $\tau_{0}$ has been
correlated with the solubility of nanoparticles in explaining
fluid-crystal coexistence curves \cite{ROS,ROS1}.

The free energy of a suspension of particles cannot, of course,
depend on $B_{2}$ alone. In a recent paper \cite{PR1} we
introduced a new analytical theory for protein solutions in which
the real fluid is replaced by a suspension of spheres with an
appropriately chosen adhesion of the Baxter type. The stickiness
parameter $\tau$ in the latter is computed by a variational
principle for the free energy. In our optimized Baxter model,
$\tau$ is not at all identical to $\tau_{0}$ but depends not only
on the ionic strength but also on the protein concentration. In
Ref. \cite{ROS}, Rosenbaum et al. plotted $\tau_{0}$
logarithmically as a function of the nanoparticle concentration
which effectively coarse-grains the experimental data they show.
If we zoom in on their curve, there is a lot of fine detail which
we here argue to be related to the fact that $\tau$ is a better
similarity parameter. In particular, we seek to understand the
ionic-strength dependence of the fluid-crystal coexistence curves
by going beyond theory based solely on $\tau_{0}$.

We have recently tested the optimized Baxter model on a system of
spheres interacting via an attractive Yukawa potential analyzed by
computer simulations \cite{PR2}. The stickiness parameter $\tau$,
evaluated by optimizing the free energy, is indeed a useful
similarity variable for gaining insight into the pressures and
chemical potentials from the simulations. The magnitudes of these
quantities are also well predicted by the optimized Baxter model.

However, here we will not focus on the variable $\tau$ and the fluid
phase but rather on the coexistence itself. The systems we study are
assumed to have a short enough range so that the coexistence between
two fluid phases is circumvented. An a priori theory is problematic
because we would need a quantitative theory of the crystal phase in
terms of postulated attractive forces which are currently unknown.
Theoretical efforts exist in the literature \cite{HA2,HA3,CUR} at
the expense of introducing unknown parameters which we want to avoid
here.

In practice, it may be very difficult to achieve ideal
thermodynamic equilibrium between the liquid phase and some
crystalline state. Equilibrium may not have been reached, the
crystal could be heterogeneous and the formation of aggregates
could complicate the attainment of equilibrium (see, for instance,
the discussion by Cacioppo and Pusey on lysozyme \cite{CAC}).
Nevertheless, it may still be useful to assume equilibrium is
ideally attained provided our goal is sufficiently modest. The
balance of chemical potentials has been used before to acquire
information about the crystal from the solubility in the fluid
phase \cite{GUO}. Our concern here will be to try to gain insight
into the ionic-strength dependence of the thermodynamic properties
of the crystal. We may argue that this dependence could be
approximated by a Donnan equilibrium so it would not be very
sensitive to the precise crystal habit adopted. We therefore
compute the protein chemical potential and osmotic pressure of the
coexisting liquid phase at the experimentally determined
solubility with the help of the optimized Baxter model. We then
investigate whether their dependence on the electrolyte
concentration agrees with that predicted by a simple crystal
model.

\section*{II. OPTIMIZED BAXTER MODEL}

We first discuss how we obtain the bare adhesion parameters via
the second virial coefficient, and then summarize the optimized
Baxter model which is a liquid state theory at finite
concentrations \cite{PR1}. We consider a system of charged
nanometer-sized particles (e.g. proteins or nanocolloids) in water
with added monovalent salt of ionic strength $I$. We suppose the
particles are spherical with radius $a$. The charge is distributed
uniformly on the particle's surface. For convenience, all
distances in this section will be scaled by the radius $a$ and all
energies by $k_{B}T$, where $k_{B}$ is Boltzmann's constant and
$T$ is the temperature. Because monovalent ions (counterions and
salt ions) are present in solution, the Coulomb repulsion between
the particles will be screened and it is here given by a far-field
Debye-H\"{u}ckel potential \cite{PR1}. The effective number
$Z_{eff}$ of charges on the sphere (taken to be positive) will
here be computed in the Poisson-Boltzmann approximation. We let
the attraction between two particles be of range much shorter than
their radius, and we model it by a potential well of depth $U_{A}$
and width $\delta\ll1$. The total interaction $U(x)$ between two
particles whose centers of mass are separated by an actual
distance $r$ is thus of the form
\begin{equation}
U(x)=\left\{
\begin{array}
[c]{ll}
\infty & \hspace{15pt} 0\leq x<2\\
U_{DH}(x)-U_{A} & \hspace{15pt} 2\leq x<2+\delta\\
U_{DH}(x) & \hspace{15pt}  x\geq2+\delta
\end{array}
\right., \label{pot}
\end{equation}
\[
x\equiv\frac{r}{a},
\]
with Debye-H\"{u}ckel interaction
\begin{equation}
U_{DH}(x)=2\xi\frac{e^{-\omega(x-2)}}{x}. \label{debye}
\end{equation}
Here, $\xi\equiv\frac{Q}{2a}\left( \frac{Z_{eff}}{1+\omega}\right)
^{2}$ and $\omega\equiv\kappa a$, which are given in terms of the
Debye length $\kappa^{-1}$ defined by $\kappa^{2}=8\pi QI$ and the
Bjerrum length $Q=q^{2}/\epsilon k_{B}T$, which equals $0.71$ nm
in water at $298$ K ($\epsilon$ is the permittivity of water, $q$
is the elementary charge); $\omega=3.28a\sqrt{I}$, if the radius
$a$ is given in nm and the ionic strength $I$ in M. We suppose 1-1
electrolyte has been added in excess so $I$ is the concentration
of added salt. We have derived the effective charge $qZ_{eff}$ in
the Poisson-Boltzmann approximation \cite{PR1}
\begin{equation}
Z_{eff}=Z-\frac{\omega^{2}}{6}\left(  \frac{Q}{a}\right)^{2}
\left( \frac{Z}{1+\omega }\right)^{3}e^{3\omega}E_{1}(3\omega).
\label{Zeff}
\end{equation}
Here $E_{1}(x)$ is the exponential integral defined by
$E_{1}(x)=\int _{x}^{\infty}dt\,t^{-1}e^{-t}$ and $qZ$ is the
actual charge per particle.

We suppose that the bare charge on the particles as a function of
the ionic strength is known from experiment, so the only unknown
parameters are $U_{A}$ and $\delta$ which are chosen to be
independent of $I$. The latter are determined by fitting
preferably complete experimental data of the second virial
coefficient $B_{2}$ as a function of the ionic strength $I$ at
constant pH to $B_{2}$ computed numerically with the help of the
expression
\begin{equation}
B_{2}=2\pi a^{3}\int_{0}^{\infty}x^{2}dx\left(1-e^{-U(x)}\right)
\end{equation}
using Eq. (\ref{pot}). We have previously done this for a wide
variety of $B_{2}$ data on lysozyme at two values of the pH (4.5
and 7.5) and we were able to obtain very good fits \cite{PR1} (see
e.g. Fig. 1 which is discussed in section III.B).

It is important to stress that though there are two adjustable
parameters $\delta$ and $U_{A}$, the actual fit in practice
depends almost solely on adjusting the single combination $\delta
\exp U_{A}$. This is because a convenient analytical approximation
of the second virial turns out to have the form \cite{PR1}
\begin{equation}
\frac{B_{2}}{B_{2}^{HS}}\simeq1+\frac{3\xi}{2\omega}-\frac{3}{2}e^{-\xi}\delta
e^{U_{A}} \label{benb2}
\end{equation}
and is able to describe the experimental data on lysozyme quite
well with an appropriate value of $\delta \exp U_{A}$. Here,
$B_{2}^{HS}$ is the second virial coefficient pertaining to hard
spheres. We note that Eq. (\ref{benb2}) disagrees starkly with an
approximation put forward earlier \cite{WAR}, both with regard to
the pure electrostatic and the adhesive contributions. In
particular, the third i.e. adhesion term in Eq. (\ref{benb2}) is
not at all independent of the ionic strength but rather diminishes
fast as the electrolyte concentration is lowered. Furthermore, the
pure electrostatic term cannot be derived from a Donnan
equilibrium as we point out in Section IV.

At high salt concentrations, the parameter $\xi$ becomes small
owing to screening so $B_{2}$ becomes lower than the hard sphere
value as can be seen from Eq. (\ref{benb2}). Nevertheless, the
electrostatic repulsion still exerts itself, so an effective
adhesion parameter we may wish to introduce would be smaller than
the bare value. We therefore adopt a similar strategy to the
liquid state at finite concentrations by first introducing a
suitable reference state amenable to analytical computation
\cite{PR1}. This is a solution of hard spheres whose radius is
still $a$ but with a Baxter adhesion potential whose strength is
defined by a suitable stickiness parameter $\tau$. The statistical
properties of this suspension as a function of the volume fraction
of spheres $\eta$ ($=4\pi a^{3}/3$ times number density) may be
solved in the Percus-Yevick approximation \cite{BAX}. The
parameter $\tau$ is adjustable and is computed via a variational
principle for the free energy. The latter may be written as a
functional expansion in terms of the so-called blip function which
is the difference in Mayer functions of the respective
interactions (Eq. (\ref{pot}) and the Baxter interaction)
\cite{PR1,AND}. We set the first-order deviation from the free
energy pertaining to the reference state equal to zero. This
determines $\tau$ which depends not only on the well parameters
$\delta$ and $U_{A}$ and electrostatic variables $\omega$ and
$\xi$ but also on the volume fraction of nanospheres. It is given
by \cite{PR1}
\begin{equation}
\frac{1}{\tau}=3\epsilon\left[  \left( e^{U_{A}}e^{-\frac{\xi
}{1+\epsilon/2}e^{-\omega\epsilon}}-1\right) \left(  1+\left(
1+H\right) \epsilon\right)  +\left(  e^{U_{A}}e^{-\xi}-1\right)
\right],\label{valt}
\end{equation}
where
\begin{equation}
\epsilon=\delta-K^{-1}\left[(1+\delta H)P_{1}+HP_{2}\right],
\label{vale}
\end{equation}
\begin{equation}
P_{1}=\frac{8}{\omega^{2}}(1+\omega\delta)M+\frac{16}{\omega}
\left(\frac{M}{1+M}\right), \label{P1}
\end{equation}
\begin{equation}
P_{2}=\frac{8}{\omega^{3}}(2+\omega\delta)M+\frac{16}{\omega^{2}}\ln(1+M),
\label{P2}
\end{equation}
\begin{equation}
M\equiv\xi e^{-\omega\delta}/4,
\end{equation}
\begin{equation}
K_{1}=2\left( e^{U_{A}}e^{-\frac{\xi}{1+\delta
/2}e^{-\omega\delta}}-1\right) \left(  1+\left(  1+H\right)
\delta\right) +2\left(
e^{U_{A}}e^{-\frac{\xi}{1+\epsilon/2}e^{-\omega\epsilon}}-1\right)
\left(  1+\left(  1+H\right)  \epsilon\right), \label{K1}
\end{equation}
\begin{equation}
H=\frac{\eta}{2\tau(1-\eta)}\left(
\frac{\eta(1-\eta)}{12}\lambda^{2}
-\frac{1+11\eta}{12}\lambda+\frac{1+5\eta}{1-\eta}-\frac{9(1+\eta)}
{2(1-\eta)^{2}}\frac{1}{\lambda}\right) \label{H}
\end{equation}
and $\lambda$ is given by
\begin{equation}
\tau=\frac{1+\eta/2}{(1-\eta)^{2}}\frac{1}{\lambda}-\frac{\eta}{1-\eta}
+\frac{\eta}{12}\lambda. \label{lambda}
\end{equation}
Note that $\tau$ is readily obtained by iteration. One starts with
initial values for $\tau$ and $\epsilon$ and then calculates
$\lambda$, $H$ and $K_{1}$ from Eqs. (\ref{K1})--(\ref{lambda}).
Then, a new value of $\epsilon$ at fixed $H$ is computed
iteratively with the help of Eqs. (\ref{vale}) and (\ref{K1}).
Next, a new value of $\tau$ is given by Eq. (\ref{valt}) and then
the cycle is repeated until the variables become stationary.

Having obtained the effective adhesion parameter $\tau$, we simply
calculate thermodynamic properties of the reference state within
the Percus-Yevick approximation. The free energy of the actual
system does deviate slightly from that of the reference state but
we have shown that the deviations are very small \cite{PR1}. To
compute the osmotic pressure $\Pi$ we use the result from the
compressibility route \cite{BAX}
\begin{equation}
\frac{\Pi
v_{0}}{k_{B}T}=\frac{\eta(1+\eta+\eta^{2})}{(1-\eta)^{3}}-
\frac{\eta^{2}(1+\eta/2)}{(1-\eta)^2}\lambda+
\frac{\eta^{3}}{36}\lambda^{3}. \label{pressure}
\end{equation}
When the roots of Eq. (\ref{lambda}) are complex, the pressure
cannot be determined for the physical realization of the liquid
state breaks down, at least with the Percus-Yevick approximation.
The chemical potential $\mu$ of the spherical particles is
determined by using the pressure from Eq. (\ref{pressure}) and the
Gibbs-Duhem equation at constant temperature \cite{BAR}
\begin{equation}
\frac{\mu-\mu_{0}}{k_{B}T}=\ln\frac{\eta}{1-\eta}+
\frac{3\eta(4-\eta)}{2(1-\eta)^2}+\frac{\Pi v_{0}}{k_{B}T}+J.
\label{chempot}
\end{equation}
Here
\begin{align}
J & = &  \frac{3}{2}\eta^{2}\lambda^{2}-
\frac{3\eta(1+4\eta)}{(1-\eta)}\lambda
+\frac{6\eta(2+\eta)}{(1-\eta)^{2}}
-\frac{18\eta}{1-\eta}\tau-\frac{6(\tau-\tau_{c})^{2}}{\tau_{c}(1-6\tau_{c})}
\ln\left|\frac{\lambda(1-\eta)-\tau_{c}^{-1}}{\tau^{-1}-\tau_{c}^{-1}}\right|+
\nonumber \\
& & +\frac{6\tau_{c}(18\tau\tau_{c}-1)^{2}}{1-6\tau_{c}}
\ln\left|\frac{\lambda(1-\eta)-18\tau_{c}}{\tau^{-1}-18\tau_{c}}\right|
\end{align}
is the contribution to the chemical potential that vanishes in the
hard-sphere limit ($\tau\rightarrow\infty$) and
\begin{equation}
\frac{\mu_{0}}{k_{B}T}= \ln \frac{1}{v_{0}}\left(\frac{h^{2}}{2\pi
mk_{B}T}\right)^{3/2}
\end{equation}
where $h$ is Planck's constant and $m$ is the mass of a sphere.
The critical value of $\tau$ below which there is a range of
densities where there is no real solution of $\lambda$, is given
by
\begin{equation}
\tau_{c}=\frac{2-\sqrt{2}}{6}.
\end{equation}

\section*{III. SOLUBILITY CURVES: CHEMICAL POTENTIAL OF THE FLUID PHASES}

\subsection*{A. Method}

Since we suppose the crystal is in thermodynamic equilibrium with
the fluid, the protein chemical potentials as well as the osmotic
pressures in both phases are uniform. The chemical potential of
the counter and co-ions must also be uniform but we will address
this issue later within a Donnan equilibrium. Solubility data from
experiment represent the particle concentration in the fluid phase
as a function of the pH and the salt concentration. Thus we
compute the chemical potential and the osmotic pressure of the
solution with the help of the optimized Baxter model of the
previous section. We have done this in two cases of nanoparticles
where we have sufficient experimental data on the second virial
coefficient to evaluate the well parameters $U_{A}$ and $\delta$
with sufficient accuracy.

\subsection*{B. Lysozyme}

The protein hen-egg-white lysozyme has been well characterized in
aqueous solutions of simple electrolytes. We here choose the
effective radius $a$ such that the volume of the model sphere is
equal to the volume of a lysozyme molecule in the tetragonal crystal
state. The latter is determined from the water content of the
tetragonal crystal (0.335 mass fraction \cite{STE}), the crystal
volume per protein molecule (29.7 nm$^3$ \cite{NAD}), the density of
the crystal (1.242 $10^{3}$ kg m$^{-3}$ \cite{STE}) and the density
of water (0.998 $10^{3}$ kg m$^{-3}$). Thus we have $a=1.61$ nm and
note that this is about 0.1 nm less than the value of 1.7 nm we used
previously \cite{PR1}, which was based on approximating the protein
by an ellipsoid of dimensions $4.5\times 3.0\times 3.0$ nm
\cite{MIK}. For the sake of consistency we here use the single value
$a=1.61$ nm in computations pertaining to both phases.

The experimental data for the second virial coefficient of
lysozyme have been discussed by us at length previously \cite{PR1}
and are presented in Fig. 1. For details on determining the
parameters $U_{A}$ and $\delta$ of the attractive potential we
also refer to Ref. \cite{PR1}. Since we are using a smaller
effective radius here, we deduce the values $U_{A}=2.89$ and
$\delta=0.182$ which are somewhat different from those derived
earlier \cite{PR1}. The values of the bare charge $qZ$ of a
lysozyme molecule as a function of the ionic strength are the same
as those used in Ref. \cite{PR1} i.e. they are determined by
interpolation from hydrogen-ion titration data in KCl \cite{KUE}.
We assume that KCl and NaCl (see below) behave identically in an
electrostatic sense. The effective charge does differ slightly
because it is a function of $a$ (see Eq. (\ref{Zeff})). We again
use the lowered effective charge $\overline{Z}=Z_{eff}-1$ instead
of the effective charge $Z_{eff}$ in order to fit $B_{2}$
accurately at lower ionic strengths when it is dominated by
electrostatics. We set $U_{A}$ and $\delta$ to be independent of
the pH.

Accurate data on the solubility $S$ as a function of the NaCl
concentration have been obtained by Cacioppo and Pusey \cite{CAC}
using column beds of tetragonal microcrystallites of lysozyme in a
range of pH and temperatures. We here employ their data at 298 K
and at three representative values of the pH. (See Tables I, II
and III.) The ionic strength $I$ in M is determined from the ionic
strength in \%w/v by the relation $I$(M)$=0.06+0.171$ $I$(\%w/v).
Here, the value $0.06$ accounts for the effective ionic strength
of the 0.1 M sodium acetate buffer used and $0.171=10/M_{NaCl}$
where $M_{NaCl}=58.44$ g mol$^{-1}$ is the molar mass of NaCl. The
dimensionless parameter $\omega$ is then given by
$\omega=5.28\sqrt{I}$, where $I$ is given in M, and
$\xi=0.220(\overline{Z}/(1+\omega))^{2}$.

The volume fraction $\eta$ of protein in the liquid phase is given
by $\eta=SN_{A}v_{0}/M$, where $N_{A}$ is Avogadro's number,
$v_{0}=4\pi a^{3}/3$ is the volume of a lysozyme molecule and
$M=14.3$ kg mol$^{-1}$ \cite{RIE} is the molar mass of lysozyme.
The parameter $\tau$ describing the effective adhesion is
determined as described in section II (see Eqs.
(\ref{valt})--(\ref{lambda})), using the values $U_{A}=2.89$ and
$\delta=0.182$. Then, the dimensionless chemical potential
$(\mu-\mu_{0})/k_{B}T$ and the dimensionless osmotic pressure $\Pi
v_{0}/k_{B}T$ are determined from Eqs. \ref{chempot} and
\ref{pressure} respectively. (See Tables I, II and III). Fig. 2
shows the chemical potential as a function of the ionic strength
$I$ at three different values of the pH. Fig. 3 shows the osmotic
pressure under the same conditions.

\subsection*{C. STA}

The next system we consider is silicotungstate (STA) in water with
three different kinds of added salt: NaCl, HCl and LiCl. STA
molecules are spherical, more or less, (see Fig. 2 in Ref.
\cite{GE}) with an effective diameter of 1.1 nm \cite{BAK,KUR}, so
we set $a=0.55$ nm. The structural formula for the polyanion
SiW$_{12}$O$_{40}^{4-}$ implies a molar mass $M_{STA}=2874.3$ g
mol$^{-1}$. We assume that the pH is low enough for the molecule
to be fully dissociated, i.e. $Z=4$.

We determine the well parameters $U_{A}$ and $\delta$ for the
attractive interaction by fitting experimental data of the second
virial coefficient in the same way as was done for lysozyme
\cite{PR1}, except we now do not adjust $Z_{eff}$. The second
virial coefficients for Li$_{4}$STA, H$_{4}$STA and Na$_{4}$STA
are taken from Ref. \cite{ZUK} and plotted in Fig. 4. In each
case, the added salt is XCl, where X represents the counterion of
the crystal. The values of the dimensionless parameters
$\omega=1.80\sqrt{I}$, $\xi=0.645(Z_{eff}/(1+\omega))^{2}$, $Z$
and $Z_{eff}$ pertaining to the data in Fig. 4 are given in Table
IV. We have set $\delta=0.05$. A least-squares fit to the data
represented in Fig. 4 then gives $U_{A}=3.30$. In fact, there is a
range of combinations of $\delta$ and $U_{A}$ that yield almost
the identical curve as long as $\delta\exp U_{A}\approx 1.36$ and
$\delta\ll 1$, so our choice of $\delta=0.05$ is a bit arbitrary.
This similarity with respect to the sole parameter $\delta\exp
U_{A}$ is in accord with our approximation for $B_{2}$ give by Eq.
(\ref{benb2}).

The solubilities for Li$_{4}$STA, H$_{4}$STA and Na$_{4}$STA have
been measured by Zukoski et al. \cite{ZUK}, where the same
electrolytes are used as in the measurements of $B_{2}$. (See
Tables V, VI and VII.) The volume fraction $\eta$ of STA is given
by $\eta=SN_{A}v_{0}/M_{X_{4}STA}$, where S is the solubility of
STA (note that here it is given in g/ml, whereas for lysozyme it
was given in g/l), $v_{0}=4\pi a^{3}/3$ is the volume of an STA
molecule and $M_{X_{4}STA}$ is the molar mass, where X again
represents the counterion in the respective cases. We have
$M_{H_{4}STA}=2878.3$ g mol$^{-1}$, $M_{Li_{4}STA}=2902.0$ g
mol$^{-1}$ and $M_{Na_{4}STA}=2966.2$ g mol$^{-1}$. The stickiness
parameter $\tau$ is determined by the method described in section
II (see Eqs. (\ref{valt})--(\ref{lambda})), using the values
$U_{A}=3.30$ and $\delta=0.05$. The chemical potential and the
osmotic pressure are again determined from Eqs. \ref{chempot} and
\ref{pressure} respectively. (See Tables V, VI and VII). We
display these thermodynamic variables as a function of the ionic
strength in Figs. 5 and 6.

\section*{IV. CRYSTAL MODEL: DONNAN EFFECT}

Having computed the thermodynamic properties of the fluid phases
of lysozyme and STA, and hence those of the respective crystal
phases under the assumption of equilibrium of the two phases, we
now attempt to gain insight into them by introducing a simple
model for the crystal. In the latter the spherical particles
either touch or are very close. There are thus minute "surfaces of
interaction" where the forces between two nearby spheres are
predominantly attractive. It is therefore reasonable to write the
thermodynamic potential $\Omega$ of a crystal of $N$ spheres in a
volume $V$ as a superposition of attractive and electrostatic
contributions to a first approximation
\begin{equation}
\Omega=\frac{1}{2}\hspace{2pt}k(c)\frac{(V-V_{0})^{2}}{V_{0}}+Nf_{el}(c,I_{c})+\Pi(S,I)V-\mu(S,I)N
\label{Omega}
\end{equation}
The crystal is immersed in a large reservoir at a constant osmotic
pressure $\Pi$ and chemical potential $\mu$ containing a saturated
solution of nanospheres at a solubility $S$ and ionic strength $I$
($\Pi$ and $\mu$ are given by Eqs. (\ref{pressure}) and
(\ref{chempot}) respectively). The crystal has elastic properties
denoted by the modulus $k$ which depends on the density $c=N/V$
and the crystal would have a volume $V_{0}$ in the absence of
electrostatic forces ($|V-V_{0}|\ll V_{0}$). Actually, the form of
the elastic energy is more complicated and depends on the precise
crystal habit \cite{LAN} but the simple harmonic form in Eq.
(\ref{Omega}) suffices for our purposes. There is a Donnan
equilibrium (see below) which leads to a salt concentration
$I_{c}$ within the interstitial region in the crystal. We adopt a
continuum approximation: the electrostatic free energy $Nf_{el}$
is computed for a lattice of charged spheres embedded in a solvent
of uniform permittivity $\epsilon$ and electrolyte concentration
$I_{c}$.

At equilibrium, $\Omega$ must be minimized
($\partial\Omega/\partial V=0$; $\partial\Omega/\partial N=0$) so
that
\begin{equation}
\Pi\simeq\Pi_{el}-k\left(\frac{V-V_{0}}{V_{0}}\right)+
\frac{1}{2}\hspace{2pt}c\hspace{2pt}\frac{\text{d}k}{\text{d}c}\left(\frac{V-V_{0}}{V_{0}}\right)^{2}
\label{Pic}
\end{equation}
\begin{equation}
\mu\simeq\mu_{el}+
\frac{1}{2}\frac{\text{d}k}{\text{d}c}\left(\frac{V-V_{0}}{V_{0}}\right)^{2}
\label{muc}
\end{equation}
We have introduced the electrostatic counterparts of the osmotic
pressure and the chemical potential of a charged sphere in the
crystal phase on the right hand sides of Eqs. (\ref{Pic}) and
(\ref{muc}). In Eq. (\ref{Pic}) the elastic term proportional to
$k$ may easily be of order $\Pi_{el}$ but the quadratic form is
negligible. In view of the fact that $\Pi_{el}=O(c\mu_{el})$, we
then have $\mu\simeq\mu_{el}$ to a good approximation from Eq.
(\ref{muc}). In effect, as we change the ionic strength of the
fluid phase, the solubility $S$ and the salt concentration $I_{c}$
within the crystal readjust themselves whereas the volume $V$
remains virtually constant. The chemical potential is modified
only by virtue of the change in electrostatic shielding about a
sphere in the lattice. But a substantial hydrostatic pressure may
be exerted within the crystal as we decrease its volume a bit.

Next, we compute the electrostatic properties of the crystal. The
colligative properties of salt-free polyelectrolytes are often
addressed in terms of a cell model in which a test cylinder is
surrounded by a boundary of similar symmetry on which the electric
field vanishes \cite{OOS}. The boundary effectively replaces the
effect of the surrounding particles on the test particle. This
picture is reasonable at low volume fractions but must break down
at high concentrations when the electric field is highly
heterogeneous. In the latter case, one of us has advocated
focusing on the voidlike regions instead of on a test particle
(see Ref. \cite{OD1} which deals with a hexagonal lattice of DNA
at very high concentrations). Thus, in a crystal of spheres we may
distinguish very small regions between particles that almost touch
which we view as thin boundary layers, and larger voids which we
will simply approximate by spheres. (We are here concerned with
spheres of high charge density which leads to counterions being
"condensed". At low charge densities, it is possible to give a
more general analysis; see the Appendix). Discrete charge effects
should prevail when evaluating the electrostatics of the boundary
layers. These energies are here assumed to be independent of the
ionic strength since the relevant scales in the boundary layers
are very small in crystals of nanoparticles.

We therefore first solve the Poisson-Boltzmann equation for a
charged void or spherical cavity of radius $b$ without salt and
then discuss the effect of monovalent salt via a Donnan
equilibrium. The charge density on the surface of the cavity is
uniform and the total number of charges is $Z$. In view of
electroneutrality there are $Z$ counterions in the cavity, each
bearing charge $-q$. Within a mean-field analysis, the counterion
density $\rho(r)$ inside the cavity is given by a Boltzmann
distribution in terms of the electrostatic potential $\Psi(r)$ at
a distance $r$ from its center
\begin{equation}
\rho(r)=\overline{\rho}\hspace{2pt}e^{q\Psi/k_{B}T}.\label{rho}
\end{equation}
We choose $\Psi=0$ at the center of the cavity so that
$\overline{\rho}$ is the actual charge density there. The charge
density $-q\rho$ is also related to $\Psi$ by Poisson's equation
\begin{equation}
\Delta\Psi=\frac{4\pi q\rho}{\epsilon}\label{poisson}
\end{equation}
leading to the Poisson-Boltzmann equation \cite{OOS} which we
conveniently express in the scaled form
\begin{equation}
\psi''(x)+\frac{2}{x}\psi'(x)=e^{\psi}. \label{PB}
\end{equation}
Here, we have defined $\lambda^{-2}\equiv 4\pi Q\overline{\rho}$,
$x\equiv r/\lambda$ and $\psi\equiv q\Psi/k_{B}T$ where $\lambda$
may be interpreted as a screening length. The two additional
boundary conditions are
\begin{equation}
\psi'(0)=0 \label{bnd2}
\end{equation}
owing to symmetry, and
\begin{equation}
\frac{b}{\lambda}\hspace{2pt}\psi'\left(\frac{b}{\lambda}\right)=
\Lambda\equiv\frac{QZ}{b} \label{bnd3}
\end{equation}
signifying the relation between the electric field and the charge
density at the surface of the cavity.

For small $x$, Eq. (\ref{PB}) admits a series expansion
$\psi(x)=Ax^{2}+Bx^{4}+...$ with $A=1/6$ and $B=O(1)$ independent
of the value of the dimensionless variable $\Lambda$. As $\Lambda$
tends to zero, Eq. (\ref{bnd3}) reduces to the condition of
electroneutrality. Electrostatic screening vanishes in this limit
and there are no counterions "condensed" on the surface of the
sphere. It is straightforward to solve Eq. (\ref{PB}) numerically
starting with $\psi(x)\rightarrow\frac{1}{6}x^{2}$ as
$x\rightarrow 0$. We have fitted the solution to the convenient
approximation
\begin{equation}
\psi(x)\approx-2\ln\left(1-\frac{x^{2}}{12}-\frac{x^{4}}{1440}-
\frac{x^{6}}{45330.3}\right),
\end{equation}
which is accurate to within 0.6 \% for $0\leq x\leq 3.273687$
($\psi(x)$ diverges at $x\approx 3.27368734$). This leads to an
effective charge given by
\begin{equation}
Z_{eff}\equiv\frac{4\pi
b^{3}\overline{\rho}}{3}=\frac{Z}{3\Lambda}\left(\frac{b}{\lambda}\right)^{2}.
\label{zeff}
\end{equation}
This is always less than the actual charge $Z$ which one may
interpret as a certain fraction of counterions being associated
near the surface if $\Lambda>0$. The effective charge $Z_{eff}$
tends to $Z$ as $\Lambda\rightarrow 0$ (for a general analysis of
this limit, see the Appendix).

We now wish to analyze the thermodynamic properties of the crystal
in the presence of simple salt which we do within a Donnan
approximation. At this stage it is well to recall the
incorrectness of applying Donnan arguments to a fluid of charged
colloidal particles. The probability of the double layers of two
particles interpenetrating is very small owing to Boltzmann
weighting. Hence, only the Debye-H\"uckel tails in their
interaction are important which represent effectively the
potential of mean force between the particles. In the case of
excess salt, we then use the McMillan-Mayer theory to calculate
the statistical mechanical properties of the fluid as has been
done in section II (see Eq. (\ref{debye}); this line of
argumentation goes back to Stigter \cite{STI}). The situation is
decidedly different when the particles are positionally ordered as
in a crystal. The double layers are forced to overlap in that
case. A usual (Donnan) approximation is then to suppose those
points at zero electric field are in equilibrium with the
reservoir \cite{OOS}. For the cavities in the crystal, this yields
\begin{equation}
I_{c}(\overline{\rho}+I_{c})=I^{2}
\end{equation}
in view of the equality of the chemical potentials of the small
ions in the respective phases. The osmotic pressure is given by
the additivity rule as argued by Oosawa for polyions within
conventional cell models \cite{OOS}
\begin{equation}
\begin{array}
[c]{ccl}
\Pi & = & (\overline{\rho}+2I_{c}-2I)k_{B}T \\
& = & \overline{\rho}\hspace{2pt}
k_{B}T\left[\sqrt{1+w^{2}}-w\right] \label{Pimod}
\end{array}
\end{equation}
\[
w\equiv\frac{2I}{\overline{\rho}}.
\]
The ions have been considered as ideal and the electrostatic
stress is zero in Eq. (\ref{Pic}). The chemical potential of the
charged cavity, accurate to the same level of approximation, is
readily computed from Eq. (\ref{Pimod}) (this is analogous to
similar calculations for cell models of long charged rods
\cite{OD2})
\begin{equation}
\mu=\mu_{ref}+\frac{Z_{eff}k_{B}T}{2}
\ln\left[\frac{\sqrt{1+w^{2}}+1}{\sqrt{1+w^{2}}-1}\right]\label{mumod}
\end{equation}
where $\mu_{ref}$ is a reference chemical potential independent of
the concentration of salt, and not identical with $\mu_{0}$ of
section II. Because the number of particles in the crystal is
equal to the number of cavities, Eq. (\ref{mumod}) also represents
the chemical potential of a charged sphere carrying $Z$ charges
but with a different $\mu_{ref}$.

\subsection*{A. Comparison with experiment}

\subsubsection*{1. Lysozyme crystal}

The volume per lysozyme molecule in the tetragonal crystal is 29.7
nm$^{3}$ (see section III.B). The radius of the effective sphere
is 1.61 nm so the volume of a cavity is 12.2 nm$^{3}$ and $b=1.43$
nm. In Tables VIII-X, we show values of $Z$ as a function of the
ionic strength $I$ at three values of the pH. From these we
calculate the dimensionless quantities $\Lambda$, $b/\lambda$ and
$Z_{eff}$ via the Poisson-Boltzmann equation. Then the pressure
and the chemical potential are evaluated using Eqs. (\ref{Pimod})
and (\ref{mumod}). (See Tables VIII-X). The curves in Fig. 7
represent the chemical potential computed in this manner together
with the predictions from the theory of the liquid state as
displayed in Fig.2. The former have been shifted by an amount
which is unknown in the present theory.

\subsubsection*{2. STA crystals}

In order to compute the chemical potential we first need to
discuss the crystal habits of STA. It is known that H$_{4}$STA is
fully dissociated for a pH larger than 5 \cite{KRO}. Zukoski et
al. \cite{ZUK,ZAM,ROS} fail to mention the pH at which their
measurements were performed, though they did deduce that all forms
of STA are dissociated in their experiments judging from the
conductivities of their solutions.

\subsubsection*{a. H$_{4}$STA.31H$_{2}$O}

This crystallizes at room temperature \cite{WYR} in the tetragonal
form (long axis $=1.856$ nm, short axes $=1.301$ nm \cite{EVA};
there are 2 STA molecules per unit cell of 3.142 nm$^{3}$). We
have earlier set the radius of an STA ion equal to 0.55 nm (see
section III.C) so the volume of H$_{2}$O per STA molecule is 0.874
nm$^{3}$ or $b=0.593$ nm. In Refs. \cite{ROS} and \cite{ZAM} the
water content of this crystal is given in terms of the molecular
formula H$_{4}$STA.31H$_{2}$O.

\subsubsection*{b. Li$_{4}$STA.24H$_{2}$O/Li$_{4}$STA.26H$_{2}$O}

Kraus describes two forms of Li$_{4}$STA with 24 H$_{2}$O and 26
H$_{2}$O molecules attached respectively \cite{KRA}. Both crystals
are rhombohedral (short axes in both cases $=1.559$ nm, long axis
$=3.898$ nm in the former, long axis $=4.118$ nm in the latter;
the angle between the short axes $=120^{\circ}$ \cite{EVA}). Kraus
also mentions that one Li ion should probably be replaced by one H
ion. There are actually three different numbers quoted for the
water content of Li$_{4}$STA.$n$H$_{2}$O in Refs.
\cite{ZUK,ZAM,ROS}: $n=21,24$ and 26! We have opted for $n=25$,
namely the average number for the crystal habits generally
accepted. As there are 6 STA molecules per unit cell, the volume
of crystal per STA molecule is 1.406 nm$^{3}$ and the radius of
our effective cavity is $b=0.553$ nm.

\subsubsection*{c. Na$_{4}$STA.18H$_{2}$O}

This crystallizes in the triclinic form within a narrow range
around 308 K \cite{WYR}. The absolute dimensions of the unit cell
do not seem to be known. We thus estimate the amount of H$_{2}$O
per STA molecule via the molecular formulas. The water content in
Na$_{4}$STA.$n$H$_{2}$O is stated to be $n=18$ in Ref. \cite{ZAM}
and $n=14$ in Ref. \cite{ROS}. The latter value seems too low and
is possibly a misprint since $n$ should be equal to 20 according
to the usual citation \cite{GME}. Accordingly, we adopt $n=18$
here to be used in the solubility studies \cite{ZUK}. A molecule
of H$_{2}$O has a volume of 0.0285 nm$^{3}$ which is based on the
amount of H$_{2}$O in the unit cells of H$_{4}$STA.31H$_{2}$O and
Li$_{4}$STA.26H$_{2}$O. Therefore, Na$_{4}$STA.18H$_{2}$O has
0.513 nm$^{3}$ H$_{2}$O per STA molecule so we have $b=0.496$ nm.

Overall, it is not clear how much H$_{2}$O is exactly present in
the STA crystals. Fortunately, the chemical potential (Eq.
(\ref{mumod})) depends only logarithmically on this quantity so
the data compiled in Tables XI-XIII are not so sensitive to this
type of uncertainty. The predicted chemical potentials are
depicted as curves in Fig. 8 together with the computations from
our theory of the liquid state (Fig. 5).

\section*{V. DISCUSSION}

Except for a slight downward adjustment of the effective charge of
lysozyme in the fluid phase, there are essentially no adjustable
parameters in our analysis. The adhesion parameters are completely
constrained by the 2nd virial curves (Figs. 1 and 4). We predict
that the chemical potentials in the fluid and solid phases should
coincide apart from an unimportant shift in the vertical offset
because the reference potential is not known exactly for the
crystal. This appears to be almost the case for lysozyme (see Fig.
7) but there is an appreciable disparity between the respective
curves in the case of the silicotungstates (see Fig. 8).
Nevertheless, we note that the shapes of the curves are the same
which implies that the logarithmic form in Eq. (\ref{mumod})
appears to be confirmed i.e. the Donnan effect seems to apply to
crystals of charged nanoparticles. This is borne out by adjusting
$Z_{eff}$ upward somewhat for both types of crystals. We then
actually attain coincident curves (see Figs. 7 and 8). A further
implication is that the precise crystal structure is unimportant
with regard to the ionic-strength dependence of $\mu$. Eq.
(\ref{mumod}) results from approximating the cavities within the
crystals by spheres; the detailed electrostatics is independent of
the salt concentration.

The coexistence equation for the osmotic pressure yields little
information (see Eq. (\ref{Pic})) because it is unclear how to
relate the adhesive forces between spheres to the elastic
properties of the crystal. To compute the latter we need insight
in the forces between the particles at the \AA ngstr\"om level
which we do not have at present. Adhesive interactions appear to
play a minor role in the STA crystals for the fluid and crystal
pressures are quite close (compare Tables V-VII with XI-XIII). By
contrast, in lysozyme crystals the osmotic pressure due to
electrostatic forces is largely balanced by sticky interactions
between touching protein molecules.

It is wise to emphasize the shortcomings in the approximations
introduced in the electrostatic interactions. Discrete charge
effects have been disregarded entirely. At the same level of
approximation we have not addressed the electrostatics of the
minutely thin boundary layers between almost touching spheres
within the crystal phase. There are cavities at nanometer scales
and these are assumed to give rise to the ionic-strength
dependence of the free energy of the crystal. The Donnan
approximation used suffers from the same drawback as always: the
effective charge $Z_{eff}$ is posited to be independent of the
electrolyte in the crystal and thus the reservoir (the fluid phase
in our case). It would be interesting to study the fluid-crystal
coexistence of globular particles of low charge density. The
counterions in the crystal would then be essentially free (see the
Appendix) and there would be less uncertainty about the magnitude
of the electrostatic interactions.

There is another potential problem in the fluid phases of
silicotungstates. At 1 M electrolyte, the solubilities of STA are
remarkably high (see Tables V-VII). It would appear that the
counterions arising from STA should contribute to the screening on
a par with the salt ions. This is not borne out by the present
analysis, however, since there is no levelling off of the chemical
potentials in the crystal phases in Figs. 7 and 8. Nevertheless, a
liquid state theory of concentrated charged nanoparticles needs to
be developed in which the counterions are duly accounted for. We
note that the interaction between the particles is not pairwise
additive in that case.

In summary, we have provided a semi-quantitative explanation for
the ionic-strength dependence of the fluid-crystal coexistence of
suspensions of charged nanoparticles. We believe this explanation
is especially forceful because we have considered two rather
disparate types of globular particles in detail. In particular,
the solubility curves of lysozyme and silicotungstate differ
markedly, yet the curves for the chemical potentials turn out to
have the same form.

\section*{VI. APPENDIX: POISSON-BOLTZMANN EQUATION IN A CRYSTAL OR POROUS MEDIUM}

It is possible to present a general analysis of the
Poisson-Boltzmann equation for the electrostatic potential
$\Psi(\vec{r})$ at position $\vec{r}$ within the aqueous
interstitial space inside a crystal (which may be considered to be
a porous medium), under appropriate conditions. The particles in
the crystal are positively charged and simple salt is absent at
first. The potential is again related to the counterion density
$\rho(\vec{r})$ via the Poisson Eq. (\ref{poisson}). Now it is
possible to discern some point $P$ in the void between several
particles where the potential is a local minimum and where the
density is $\overline{\rho}_{P}(0)$ (see Fig. 9). Point $P$ is
chosen as the origin.

If the potential is scaled analogously as in section IV, we have
$\rho(\vec{r})=\overline{\rho}_{P}\exp\psi(\vec{r})$ (see Eq.
(\ref{rho})). Thus, the Poisson-Boltzmann equation may be written
as
\begin{equation}
\Delta\psi=\lambda_{P}^{-2}e^{\psi}\label{appsi}
\end{equation}
where the screening length $\lambda_{P}$ is given by
$\lambda_{P}^{-2}=4\pi Q\overline{\rho}_{P}$.

In general, it is difficult to address Eq. (\ref{appsi}) because
$\lambda_{P}$ is unknown. But it is possible to progress if we
suppose $|\vec{d}|\lesssim\lambda_{P}$ where $|\vec{d}|$ is the
largest vector distance between $P$ and a point on the surface of
the surrounding spheres (i.e. those belonging to a cluster
enclosing the void centered on $P$). A solution of Eq.
(\ref{appsi}) must have the form $\psi(\vec{r}/\lambda_{P})$ and
may be written as a Taylor expansion to second order
\begin{equation}
\psi(\vec{r})= \frac{1}{2}\vec{r}\vec{r}:
\left.\frac{\partial^{2}\psi}{\partial\vec{r}
\partial\vec{r}}\right|_{\vec{r}=0}\label{ansatz}
\end{equation}
if $|\vec{d}|\lesssim\lambda_{P}$. Next, we have a boundary
condition on the electric field at $\vec{r}=\vec{d}$
\begin{equation}
\vec{n}\cdot\left.\frac{\partial\psi}{\partial\vec{r}}
\right|_{\vec{r}=\vec{d}}= \vec{n}\vec{d}:
\left.\frac{\partial^{2}\psi}{\partial\vec{r}
\partial\vec{r}}\right|_{\vec{r}=0} =4\pi
k_{1}\sigma_{b}Q.\label{apbc}
\end{equation}
Here, $\sigma_{b}$ is the uniform density of charge on a sphere
and $k_{1}$ is a numerical coefficient of order unity. The effect
of an internal permittivity is disregarded. The left-hand side of
Eq. (\ref{apbc}) scales as $\lambda_{P}^{-2}$ implying that
$\overline{\rho}_{P}$ must be proportional to $\sigma_{b}$. In
view of electroneutrality we also require the average of
$\rho(\vec{r})$ to be proportional to $\sigma_{b}$. Hence, the
potential $\psi(\vec{r})$ must be very small, which is consistent
with the initial Ansatz Eq. (\ref{ansatz}). We conclude that for
small enough cavitylike voids, the density of counterions is
approximately constant so that the effective charge density is
virtually equal to the actual charge density. In that case, when
the crystal is immersed in a reservoir containing monovalent
electrolyte, Eqs. (\ref{Pimod}) and (\ref{mumod}) are valid with
$\overline{\rho}$ simply given by the concentration of counterions
in the interstitial space between the spheres; $Z_{eff}=Z$ in Eq.
(\ref{mumod}). Because $|\vec{d}|=O(a)$, we ultimately require
$ZQ/a\ll 1$ as a necessary and sufficient condition for this to
hold true. In the spherical cavity approximation introduced in
Section IV, we have $b=O(a)$ so $\Lambda\ll 1$ is effectively the
same requirement (which led to $Z_{eff}=Z$).

\section*{Tables}

\hspace{0.3in}%
\begin{tabular}
[c]{|c|c|c|c|c|c|c|c|c|c|c|c|}

\hline $I$ (\%w/v) & $I$ (M) & $\omega$ & $Z$ & $Z_{eff}$ &
$\overline{Z}$ & $\xi$ & $\tau$ & $S$ (g/l) &
$\eta$ & $\frac{\mu-\mu_{0}}{k_{B}T}$ & $\frac{\Pi v_{0}}{k_{B}T}$\\

\hline $2.0$ & $0.40$ & $3.35$ & $11.1$ & $10.93$ & $9.93$ &
$1.146$ & $0.138$ & $48.7$ & $0.0358$ & $-3.55$ & $0.0319$\\

\hline $3.0$ & $0.57$ & $4.00$ & $11.2$ & $11.07$ & $10.07$ &
$0.893$ & $0.101$ & $14.0$ & $0.0103$ & $-4.69$ & $0.0097$\\

\hline $4.0$ & $0.74$ & $4.56$ & $11.4$ & $11.29$ & $10.29$ &
$0.755$ & $0.087$ & $4.30$ & $0.00316$ & $-5.80$ & $0.0031$\\

\hline $5.0$ & $0.92$ & $5.05$ & $11.6$ & $11.51$ & $10.51$ &
$0.663$ & $0.079$ & $3.11$ & $0.00229$ & $-6.12$ & $0.0022$\\

\hline $7.0$ & $1.26$ & $5.92$ & $11.7$ & $11.63$ & $10.63$ &
$0.519$ & $0.070$ & $1.36$ & $0.00100$ & $-6.93$ & $0.0010$\\

\hline
\end{tabular}
\smallskip

\vspace{5pt}

TABLE I: The charge $Z$ of hen-egg-white lysozyme (from Ref.
\cite{KUE}), the effective charge $Z_{eff}$ (from Eq. \ref{Zeff}),
the lowered effective charge $\overline{Z}=Z_{eff}-1$, the
dimensionless interaction parameters $\omega$, $\xi$ and $\tau$,
the solubility of lysozyme $S$, the volume fraction $\eta$, the
dimensionless chemical potential $(\mu-\mu_{0})/k_{B}T$ and the
dimensionless pressure $\Pi v_{0}/k_{B}T$ as a function of the
ionic strength $I$ in the fluid phase. The pH equals 4.0 and $\xi$
has been calculated using the lowered effective charge
$\overline{Z}$.

\vspace{15pt}

\hspace{0.3in}%
\begin{tabular}
[c]{|c|c|c|c|c|c|c|c|c|c|c|c|}

\hline $I$ (\%w/v) & $I$ (M) & $\omega$ & $Z$ & $Z_{eff}$ &
$\overline{Z}$ & $\xi$ & $\tau$ & $S$ (g/l) &
$\eta$ & $\frac{\mu-\mu_{0}}{k_{B}T}$ & $\frac{\Pi v_{0}}{k_{B}T}$\\

\hline $2.0$ & $0.40$ & $3.35$ & $10.2$ & $10.07$ & $9.07$ &
$0.956$ & $0.114$ & $30.1$ & $0.0221$ & $-4.02$ & $0.0199$\\

\hline $3.0$ & $0.57$ & $4.00$ & $10.3$ & $10.20$ & $9.20$ &
$0.745$ & $0.089$ & $10.3$ & $0.00759$ & $-4.99$ & $0.0072$\\

\hline $4.0$ & $0.74$ & $4.56$ & $10.3$ & $10.22$ & $9.22$ &
$0.606$ & $0.078$ & $5.22$ & $0.00385$ & $-5.63$ & $0.0037$\\

\hline $5.0$ & $0.92$ & $5.05$ & $10.4$ & $10.33$ & $9.33$ &
$0.523$ & $0.072$ & $3.43$ & $0.00253$ & $-6.03$ & $0.0025$\\

\hline $7.0$ & $1.26$ & $5.92$ & $10.4$ & $10.35$ & $9.35$ &
$0.401$ & $0.065$ & $1.87$ & $0.00137$ & $-6.62$ & $0.0014$\\

\hline
\end{tabular}
\smallskip

\vspace{5pt}

TABLE II: Same as Table I, but now with a pH equal to 4.5.

\vspace{15pt}

\hspace{0.3in}%
\begin{tabular}
[c]{|c|c|c|c|c|c|c|c|c|c|c|c|}

\hline $I$ (\%w/v) & $I$ (M) & $\omega$ & $Z$ & $Z_{eff}$ &
$\overline{Z}$ & $\xi$ & $\tau$ & $S$ (g/l) &
$\eta$ & $\frac{\mu-\mu_{0}}{k_{B}T}$ & $\frac{\Pi v_{0}}{k_{B}T}$\\

\hline $2.0$ & $0.40$ & $3.35$ & $9.1$ & $9.00$ & $8.00$ & $0.745$
& $0.094$ & $17.6$ & $0.0130$ & $-4.52$ & $0.0119$\\

\hline $3.0$ & $0.57$ & $4.00$ & $9.1$ & $9.03$ & $8.03$ &
$0.568$ & $0.077$ & $7.38$ & $0.00543$ & $-5.31$ & $0.0052$\\

\hline $4.0$ & $0.74$ & $4.56$ & $9.2$ & $9.14$ & $8.14$ &
$0.473$ & $0.070$ & $4.72$ & $0.00347$ & $-5.73$ & $0.0034$\\

\hline $5.0$ & $0.92$ & $5.05$ & $9.2$ & $9.15$ & $8.15$ &
$0.399$ & $0.066$ & $3.63$ & $0.00267$ & $-5.98$ & $0.0026$\\

\hline $7.0$ & $1.26$ & $5.92$ & $9.1$ & $9.07$ & $8.07$ &
$0.299$ & $0.060$ & $2.46$ & $0.00181$ & $-6.36$ & $0.0018$\\

\hline
\end{tabular}
\smallskip

\vspace{5pt}

TABLE III: Same as Table I, but now with a pH equal to 5.4.

\vspace{15pt}

\hspace{0.3in}%
\begin{tabular}
[c]{|c|c|c|c|c|c|c|c|c|c|c|c|}

\hline $I$ (M) & \hspace{12pt}$\omega$\hspace{12pt} &
\hspace{10pt}$Z$\hspace{10pt} & \hspace{5pt}$Z_{eff}$\hspace{5pt}
& \hspace{15pt}$\xi\hspace{15pt}$\\

\hline $0.3$ & $0.99$ & $4.00$ & $3.42$ & $1.905$\\

\hline $1.0$ & $1.80$ & $4.00$ & $3.58$ & $1.054$\\

\hline $3.0$ & $3.13$ & $4.00$ & $3.76$ & $0.536$\\

\hline $4.0$ & $3.61$ & $4.00$ & $3.80$ & $0.439$\\

\hline $5.0$ & $4.03$ & $4.00$ & $3.83$ & $0.373$\\

\hline
\end{tabular}
\smallskip

\vspace{5pt}

TABLE IV: Values of the bare charge $Z$ of STA, the effective
charge $Z_{eff}$ (from Eq. (\ref{Zeff})) and the dimensionless
interaction parameters $\omega=1.80\sqrt{I}$ and
$\xi=0.645(Z_{eff}/(1+\omega))^{2}$ as a function of the ionic
strength $I$. These entries apply to the data plotted in Fig. 4.

\vspace{15pt}

\hspace{0.3in}%
\begin{tabular}
[c]{|c|c|c|c|c|c|c|c|c|c|c|c|}

\hline $I$ (M) & \hspace{12pt}$\omega$\hspace{12pt} &
\hspace{10pt}$Z$\hspace{10pt} & \hspace{5pt}$Z_{eff}$\hspace{5pt}
& \hspace{15pt}$\xi$\hspace{15pt} &
\hspace{15pt}$\tau$\hspace{15pt} & $S$ (g/ml) &
\hspace{15pt}$\eta$\hspace{15pt} &
\hspace{10pt}$\frac{\mu-\mu_{0}}{k_{B}T}$\hspace{10pt} &
\hspace{10pt}$\frac{\Pi v_{0}}{k_{B}T}\hspace{10pt}$\\

\hline $1.0$ & $1.80$ & $4.0$ & $3.58$ & $1.054$ & $0.786$ &
$1.94$ & $0.284$ & $1.30$ & $0.706$\\

\hline $2.0$ & $2.55$ & $4.0$ & $3.70$ & $0.700$ & $0.358$ &
$1.67$ & $0.243$ & $-0.48$ & $0.375$\\

\hline $3.0$ & $3.13$ & $4.0$ & $3.76$ & $0.536$ & $0.255$ &
$1.36$ & $0.198$ & $-1.40$ & $0.226$\\

\hline $4.0$ & $3.61$ & $4.0$ & $3.80$ & $0.439$ & $0.215$ &
$1.11$ & $0.162$ & $-1.91$ & $0.158$\\

\hline $5.0$ & $4.03$ & $4.0$ & $3.83$ & $0.373$ & $0.193$ &
$0.57$ & $0.0830$ & $-2.65$ & $0.077$\\

\hline
\end{tabular}
\smallskip

\vspace{5pt}

TABLE V: The charge $Z$ of STA, the effective charge $Z_{eff}$
(from Eq. (\ref{Zeff})), the dimensionless interaction parameters
$\omega$, $\xi$ and $\tau$, the solubility $S$ of H$_{4}$STA, the
volume fraction $\eta$, the dimensionless chemical potential
$(\mu-\mu_{0})/k_{B}T$ and the dimensionless pressure $\Pi
v_{0}/k_{B}T$ as a function of the ionic strength $I$ in the fluid
phase. Here the counterion is H$^{+}$ and the added salt is HCl.
$\xi$ has been calculated using the effective charge $Z_{eff}$.

\vspace{15pt}

\hspace{0.3in}%
\begin{tabular}
[c]{|c|c|c|c|c|c|c|c|c|c|c|c|}

\hline $I$ (M) & \hspace{12pt}$\omega$\hspace{12pt} &
\hspace{10pt}$Z$\hspace{10pt} & \hspace{5pt}$Z_{eff}$\hspace{5pt}
& \hspace{15pt}$\xi$\hspace{15pt} &
\hspace{15pt}$\tau$\hspace{15pt} & $S$ (g/ml) &
\hspace{15pt}$\eta$\hspace{15pt} &
\hspace{10pt}$\frac{\mu-\mu_{0}}{k_{B}T}$\hspace{10pt} &
\hspace{10pt}$\frac{\Pi v_{0}}{k_{B}T}\hspace{10pt}$\\

\hline $1.0$ & $1.80$ & $4.00$ & $3.58$ & $1.054$ & $0.668$ &
$2.15$ & $0.312$ & $1.55$ & $0.811$\\

\hline $2.0$ & $2.55$ & $4.00$ & $3.70$ & $0.700$ & $0.352$ &
$1.85$ & $0.267$ & $-0.26$ & $0.433$\\

\hline $3.0$ & $3.13$ & $4.00$ & $3.76$ & $0.536$ & $0.255$ &
$1.36$ & $0.197$ & $-1.41$ & $0.224$\\

\hline $4.0$ & $3.61$ & $4.00$ & $3.80$ & $0.439$ & $0.215$ &
$0.81$ & $0.118$ & $-2.23$ & $0.114$\\

\hline $5.0$ & $4.03$ & $4.00$ & $3.83$ & $0.373$ & $0.193$ &
$0.32$ & $0.0469$ & $-3.16$ & $0.045$\\

\hline
\end{tabular}
\smallskip

\vspace{5pt}

TABLE VI: Same as Table V, but now with Li$^{+}$ as counterion and
LiCl as added salt.

\vspace{15pt}

\hspace{0.3in}%
\begin{tabular}
[c]{|c|c|c|c|c|c|c|c|c|c|c|c|}

\hline $I$ (M) & \hspace{12pt}$\omega$\hspace{12pt} &
\hspace{10pt}$Z$\hspace{10pt} & \hspace{5pt}$Z_{eff}$\hspace{5pt}
& \hspace{15pt}$\xi$\hspace{15pt} &
\hspace{15pt}$\tau$\hspace{15pt} & $S$ (g/ml) &
\hspace{15pt}$\eta$\hspace{15pt} &
\hspace{10pt}$\frac{\mu-\mu_{0}}{k_{B}T}$\hspace{10pt} &
\hspace{10pt}$\frac{\Pi v_{0}}{k_{B}T}\hspace{10pt}$\\

\hline $1.0$ & $1.80$ & $4.0$ & $3.58$ & $1.054$ & $0.906$ &
$1.85$ & $0.262$ & $1.07$ & $0.625$\\

\hline $2.0$ & $2.55$ & $4.0$ & $3.70$ & $0.700$ & $0.359$ &
$1.67$ & $0.236$ & $-0.54$ & $0.358$\\

\hline $3.0$ & $3.13$ & $4.0$ & $3.76$ & $0.536$ & $0.255$ &
$1.36$ & $0.192$ & $-1.44$ & $0.218$\\

\hline $4.0$ & $3.61$ & $4.0$ & $3.80$ & $0.439$ & $0.215$ &
$0.74$ & $0.104$ & $-2.35$ & $0.100$\\

\hline $5.0$ & $4.03$ & $4.0$ & $3.83$ & $0.373$ & $0.193$ &
$0.46$ & $0.0657$ & $-2.86$ & $0.062$\\

\hline
\end{tabular}
\smallskip

\vspace{5pt}

TABLE VII: Same as Table V, but now with Na$^{+}$ as counterion
and NaCl as added salt.

\vspace{10pt}

\hspace{0.3in}
\begin{tabular}
[c]{|c|c|c|c|c|c|c|c|c|c|c|c|}

\hline $I$ (\%w/v) & $I$ (M) & \hspace{12pt}$Z$\hspace{12pt} &
\hspace{12pt}$\Lambda$\hspace{12pt} &
\hspace{12pt}$b/\lambda$\hspace{12pt} &
\hspace{12pt}$Z_{eff}$\hspace{12pt} &
\hspace{12pt}$w$\hspace{12pt} &
\hspace{12pt}$\frac{\mu}{k_{B}T}$\hspace{12pt}
& \hspace{12pt}$\frac{\Pi v_{0}}{k_{B}T}\hspace{12pt}$\\

\hline $2.0$ & $0.40$ & $11.1$ & $5.51$ & $2.52$ & $4.27$ & $1.39$ & $2.87$ & $1.97$\\

\hline $3.0$ & $0.57$ & $11.2$ & $5.56$ & $2.53$ & $4.29$ & $1.97$ & $2.10$ & $1.47$\\

\hline $4.0$ & $0.74$ & $11.4$ & $5.66$ & $2.54$ & $4.32$ & $2.54$ & $1.66$ & $1.17$\\

\hline $5.0$ & $0.92$ & $11.6$ & $5.76$ & $2.55$ & $4.35$ & $3.10$ & $1.38$ & $0.98$\\

\hline $7.0$ & $1.26$ & $11.7$ & $5.81$ & $2.55$ & $4.36$ & $4.24$ & $1.02$ & $0.72$\\

\hline
\end{tabular}
\smallskip

\vspace{5pt}

TABLE VIII: The ionic strength $I$, the actual number of charges
$Z$, the effective number $Z_{eff}$, the chemical potential $\mu$
and the osmotic pressure $\Pi$ for a lysozyme crystal at pH 4.0
(the reference chemical potential has been set equal to zero).

\vspace{15pt}

\hspace{0.3in}%
\begin{tabular}
[c]{|c|c|c|c|c|c|c|c|c|c|c|c|}

\hline $I$ (\%w/v) & $I$ (M) & \hspace{12pt}$Z$\hspace{12pt} &
\hspace{12pt}$\Lambda$\hspace{12pt} &
\hspace{12pt}$b/\lambda$\hspace{12pt} &
\hspace{12pt}$Z_{eff}$\hspace{12pt} &
\hspace{12pt}$w$\hspace{12pt} &
\hspace{12pt}$\frac{\mu}{k_{B}T}$\hspace{12pt}
& \hspace{12pt}$\frac{\Pi v_{0}}{k_{B}T}\hspace{12pt}$\\

\hline $2.0$ & $0.40$ & $10.2$ & $5.07$ & $2.48$ & $4.13$ & $1.43$ & $2.69$ & $1.86$\\

\hline $3.0$ & $0.57$ & $10.3$ & $5.12$ & $2.49$ & $4.15$ & $2.04$ & $1.96$ & $1.38$\\

\hline $4.0$ & $0.74$ & $10.3$ & $5.12$ & $2.49$ & $4.15$ & $2.64$ & $1.53$ & $1.08$\\

\hline $5.0$ & $0.92$ & $10.4$ & $5.17$ & $2.49$ & $4.16$ & $3.24$ & $1.27$ & $0.90$\\

\hline $7.0$ & $1.26$ & $10.4$ & $5.17$ & $2.49$ & $4.16$ & $4.45$ & $0.93$ & $0.66$\\

\hline
\end{tabular}
\smallskip

\vspace{5pt}

TABLE IX: Same as Table VIII, but now for pH 4.5.

\vspace{15pt}

\hspace{0.3in}%
\begin{tabular}
[c]{|c|c|c|c|c|c|c|c|c|c|c|c|}

\hline $I$ (\%w/v) & $I$ (M) & \hspace{12pt}$Z$\hspace{12pt} &
\hspace{12pt}$\Lambda$\hspace{12pt} &
\hspace{12pt}$b/\lambda$\hspace{12pt} &
\hspace{12pt}$Z_{eff}$\hspace{12pt} &
\hspace{12pt}$w$\hspace{12pt} &
\hspace{12pt}$\frac{\mu}{k_{B}T}$\hspace{12pt}
& \hspace{12pt}$\frac{\Pi v_{0}}{k_{B}T}\hspace{12pt}$\\

\hline $2.0$ & $0.40$ & $9.1$ & $4.52$ & $2.42$ & $3.93$ & $1.51$ & $2.45$ & $1.70$\\

\hline $3.0$ & $0.57$ & $9.1$ & $4.52$ & $2.42$ & $3.93$ & $2.15$ & $1.77$ & $1.25$\\

\hline $4.0$ & $0.74$ & $9.2$ & $4.57$ & $2.43$ & $3.95$ & $2.77$ & $1.40$ & $0.99$\\

\hline $5.0$ & $0.92$ & $9.2$ & $4.57$ & $2.43$ & $3.95$ & $3.41$ & $1.14$ & $0.81$\\

\hline $7.0$ & $1.26$ & $9.1$ & $4.52$ & $2.42$ & $3.93$ & $4.71$ & $0.83$ & $0.59$\\

\hline
\end{tabular}
\smallskip

\vspace{5pt}

TABLE X: Same as Table VIII, but now for pH 5.4.

\vspace{10pt}

\hspace{0.3in}
\begin{tabular}
[c]{|c|c|c|c|c|c|c|c|c|c|c|c|}

\hline $I$ (\%w/v) & $I$ (M) & \hspace{12pt}$Z$\hspace{12pt} &
\hspace{12pt}$\Lambda$\hspace{12pt} &
\hspace{12pt}$b/\lambda$\hspace{12pt} &
\hspace{12pt}$Z_{eff}$\hspace{12pt} &
\hspace{12pt}$w$\hspace{12pt} &
\hspace{12pt}$\frac{\mu}{k_{B}T}$\hspace{12pt}
& \hspace{12pt}$\frac{\Pi v_{0}}{k_{B}T}\hspace{12pt}$\\

\hline $1.0$ & $0.60$ & $4.0$ & $4.789$ & $2.452$ & $1.67$ & $0.629$ & $2.08$ & $0.738$\\

\hline $2.0$ & $1.20$ & $4.0$ & $4.789$ & $2.452$ & $1.67$ & $1.257$ & $1.22$ & $0.466$\\

\hline $3.0$ & $1.81$ & $4.0$ & $4.789$ & $2.452$ & $1.67$ & $1.886$ & $0.85$ & $0.332$\\

\hline $4.0$ & $2.41$ & $4.0$ & $4.789$ & $2.452$ & $1.67$ & $2.514$ & $0.65$ & $0.256$\\

\hline $5.0$ & $3.01$ & $4.0$ & $4.789$ & $2.452$ & $1.67$ & $3.143$ & $0.52$ & $0.207$\\

\hline
\end{tabular}
\smallskip

\vspace{5pt}

TABLE XI: Same as Table VIII but now for H$_{4}$STA.31H$_{2}$O.

\vspace{15pt}

\hspace{0.3in}%
\begin{tabular}
[c]{|c|c|c|c|c|c|c|c|c|c|c|c|}

\hline $I$ (\%w/v) & $I$ (M) & \hspace{12pt}$Z$\hspace{12pt} &
\hspace{12pt}$\Lambda$\hspace{12pt} &
\hspace{12pt}$b/\lambda$\hspace{12pt} &
\hspace{12pt}$Z_{eff}$\hspace{12pt} &
\hspace{12pt}$w$\hspace{12pt} &
\hspace{12pt}$\frac{\mu}{k_{B}T}$\hspace{12pt}
& \hspace{12pt}$\frac{\Pi v_{0}}{k_{B}T}\hspace{12pt}$\\

\hline $1.0$ & $0.60$ & $4.0$ & $5.134$ & $2.488$ & $1.61$ & $0.531$ & $2.23$ & $0.950$\\

\hline $2.0$ & $1.20$ & $4.0$ & $5.134$ & $2.488$ & $1.61$ & $1.062$ & $1.35$ & $0.626$\\

\hline $3.0$ & $1.81$ & $4.0$ & $5.134$ & $2.488$ & $1.61$ & $1.594$ & $0.95$ & $0.455$\\

\hline $4.0$ & $2.41$ & $4.0$ & $5.134$ & $2.488$ & $1.61$ & $2.125$ & $0.73$ & $0.353$\\

\hline $5.0$ & $3.01$ & $4.0$ & $5.134$ & $2.488$ & $1.61$ & $2.656$ & $0.59$ & $0.288$\\

\hline
\end{tabular}
\smallskip

\vspace{5pt}

TABLE XII: Same as Table VIII but now for Li$_{4}$STA.25H$_{2}$O.

\vspace{15pt}

\hspace{0.3in}%
\begin{tabular}
[c]{|c|c|c|c|c|c|c|c|c|c|c|c|}

\hline $I$ (\%w/v) & $I$ (M) & \hspace{12pt}$Z$\hspace{12pt} &
\hspace{12pt}$\Lambda$\hspace{12pt} &
\hspace{12pt}$b/\lambda$\hspace{12pt} &
\hspace{12pt}$Z_{eff}$\hspace{12pt} &
\hspace{12pt}$w$\hspace{12pt} &
\hspace{12pt}$\frac{\mu}{k_{B}T}$\hspace{12pt}
& \hspace{12pt}$\frac{\Pi v_{0}}{k_{B}T}\hspace{12pt}$\\

\hline $1.0$ & $0.60$ & $4.0$ & $5.721$ & $2.542$ & $1.51$ & $0.410$ & $2.45$ & $1.374$\\

\hline $2.0$ & $1.20$ & $4.0$ & $5.721$ & $2.542$ & $1.51$ & $0.819$ & $1.55$ & $0.970$\\

\hline $3.0$ & $1.81$ & $4.0$ & $5.721$ & $2.542$ & $1.51$ & $1.229$ & $1.12$ & $0.728$\\

\hline $4.0$ & $2.41$ & $4.0$ & $5.721$ & $2.542$ & $1.51$ & $1.639$ & $0.87$ & $0.575$\\

\hline $5.0$ & $3.01$ & $4.0$ & $5.721$ & $2.542$ & $1.51$ & $2.049$ & $0.71$ & $0.473$\\

\hline
\end{tabular}
\smallskip

\vspace{5pt}

TABLE XIII: Same as Table VIII but now for Na$_{4}$STA.18H$_{2}$O.

\section*{Figure Captions}

FIG. 1. The second virial coefficient of lysozyme as a function of
the ionic strength. The second virial coefficient is scaled by the
hard sphere value $B_{2}^{HS}$. The data are taken from a variety
of experiments; see Ref. \cite{PR1} for more details. The added
salt is NaCl. The solid line is a fit to the data with
$U_{A}=2.89$ and $\delta=0.182$.

FIG. 2. The dimensionless chemical potential of lysozyme in the
fluid phase as a function of the ionic strength at pH 4.0
(diamonds), pH 4.5 (squares) and pH 5.4 (triangles). See also
Tables I, II and III.

FIG. 3. The dimensionless pressure of lysozyme in the fluid phase
as a function of the ionic strength at pH 4.0 (diamonds), pH 4.5
(squares) and pH 5.4 (triangles). See also Tables I, II and III.

FIG. 4. The second virial coefficient of STA as a function of the
ionic strength. The second virial coefficient is scaled by the
hard sphere value $B_{2}^{HS}$. The experimental data are taken
from Zukoski et al. (Ref. \cite{ZUK}). The added salt is LiCl
(diamonds), HCl (squares) and NaCl (triangles) respectively and in
all cases the counterion of STA is the same as that of the salt.
The solid line is a fit to the experimental data with $U_{A}=3.30$
and $\delta=0.05$.

FIG. 5. The dimensionless chemical potential of STA in the fluid
phase as a function of the ionic strength. The added salt is LiCl
(diamonds), HCl (squares) and NaCl (triangles) respectively and in
all cases the counterion of STA is the same as that of the salt.
See also Tables V, VI and VII.

FIG. 6. The dimensionless pressure of STA in the fluid phase as a
function of the ionic strength. The added salt is LiCl (diamonds),
HCl (squares) and NaCl (triangles) respectively and in all cases
the counterion of STA is the same as that of the salt. See also
Tables V, VI and VII.

FIG. 7. The chemical potential of lysozyme in the fluid phase as a
function of the ionic strength at pH 4.0 (diamonds), pH 4.5
(squares) and pH 5.4 (triangles) (see Fig. 2). The solid lines
denote predictions from the theory of the crystalline state (Eq.
(\ref{mumod})), with the effective charge from Table VIII-X. The
shift in chemical potential in units of $k_{B}T$ has been chosen
to be 7.4 (light grey line, pH 4.0), 7.3 (black line, pH 4.5) and
7.15 (dark grey line, pH 5.4) respectively. Dashed line denotes
the theory of the crystal for $Z_{eff}=5.0$ (shift $=7.9$) and the
dash-dotted line for $Z_{eff}=5.9$ (shift $=8.7$).

FIG. 8. Chemical potential of STA in the fluid phase as a function
of the ionic strength (see Fig. 5). Salt added: LiCl (diamonds);
HCl (squares); NaCl (triangles) (counterion of STA is the same as
that of the salt). Solid lines denote predictions from the theory
of the crystalline state (Eq. (\ref{mumod})), with the effective
charge from Table XI-XIII. The shift in chemical potential in
units of $k_{B}T$ is 2.3 (black line, H$_{4}$STA.31H$_{2}$O), 2.4
(dark grey line, Li$_{4}$STA.25H$_{2}$O) and 2.55 (light grey
line, Na$_{4}$STA.18H$_{2}$O) respectively. Dashed line denotes
predictions from the theory of the crystal for $Z_{eff}=2.8$
(shift $=5.5$).

FIG. 9. Point $P$ in a void of the crystal.

\hspace{60pt}

  \begin{figure}[h]
  \begin{minipage}[t]{.45\textwidth}
    \begin{center}
      \epsfig{file=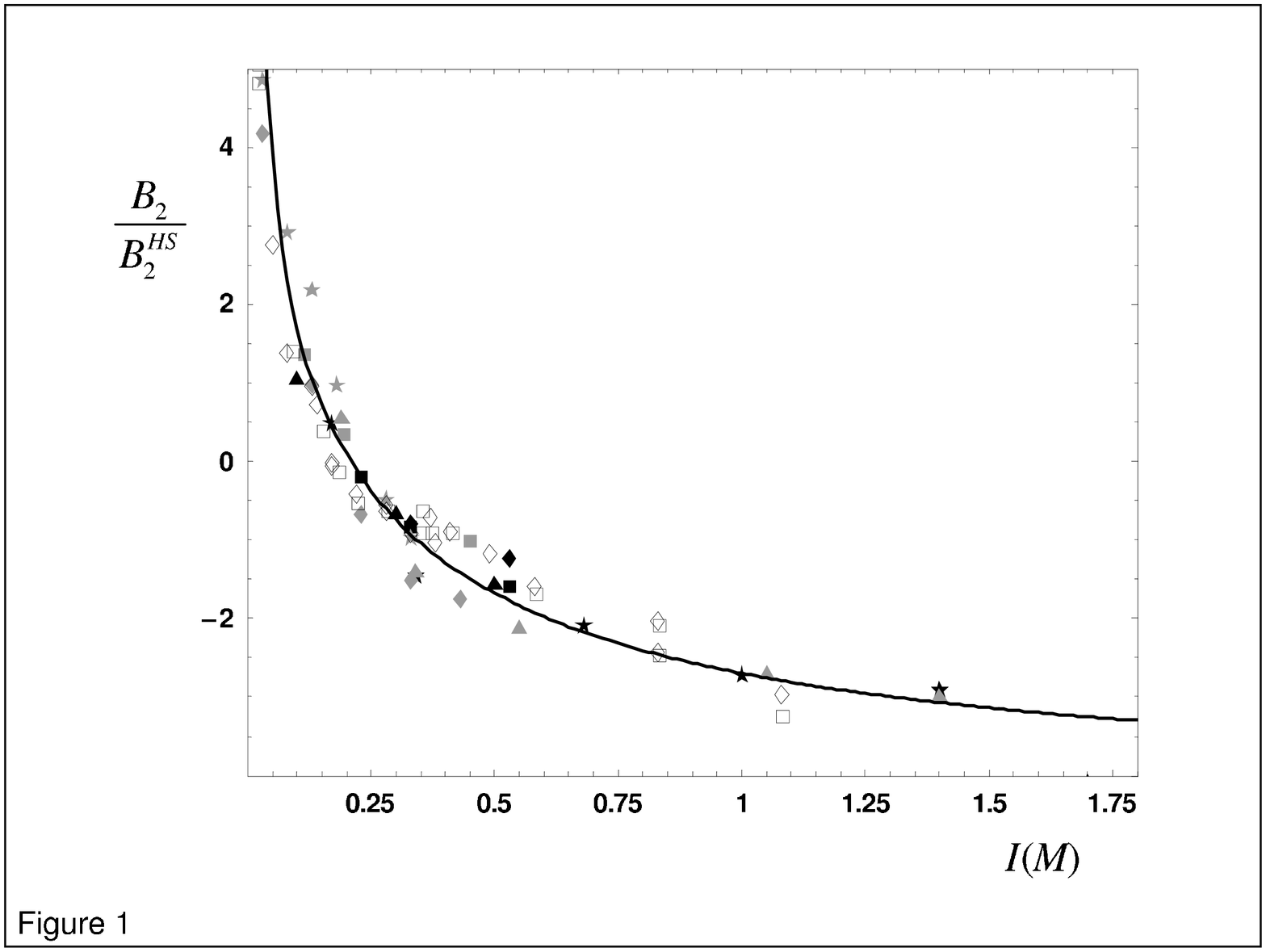, scale=0.5, angle=270, width=250pt}
    \end{center}
  \end{minipage}
  \begin{minipage}[t]{.45\textwidth}
    \begin{center}
      \epsfig{file=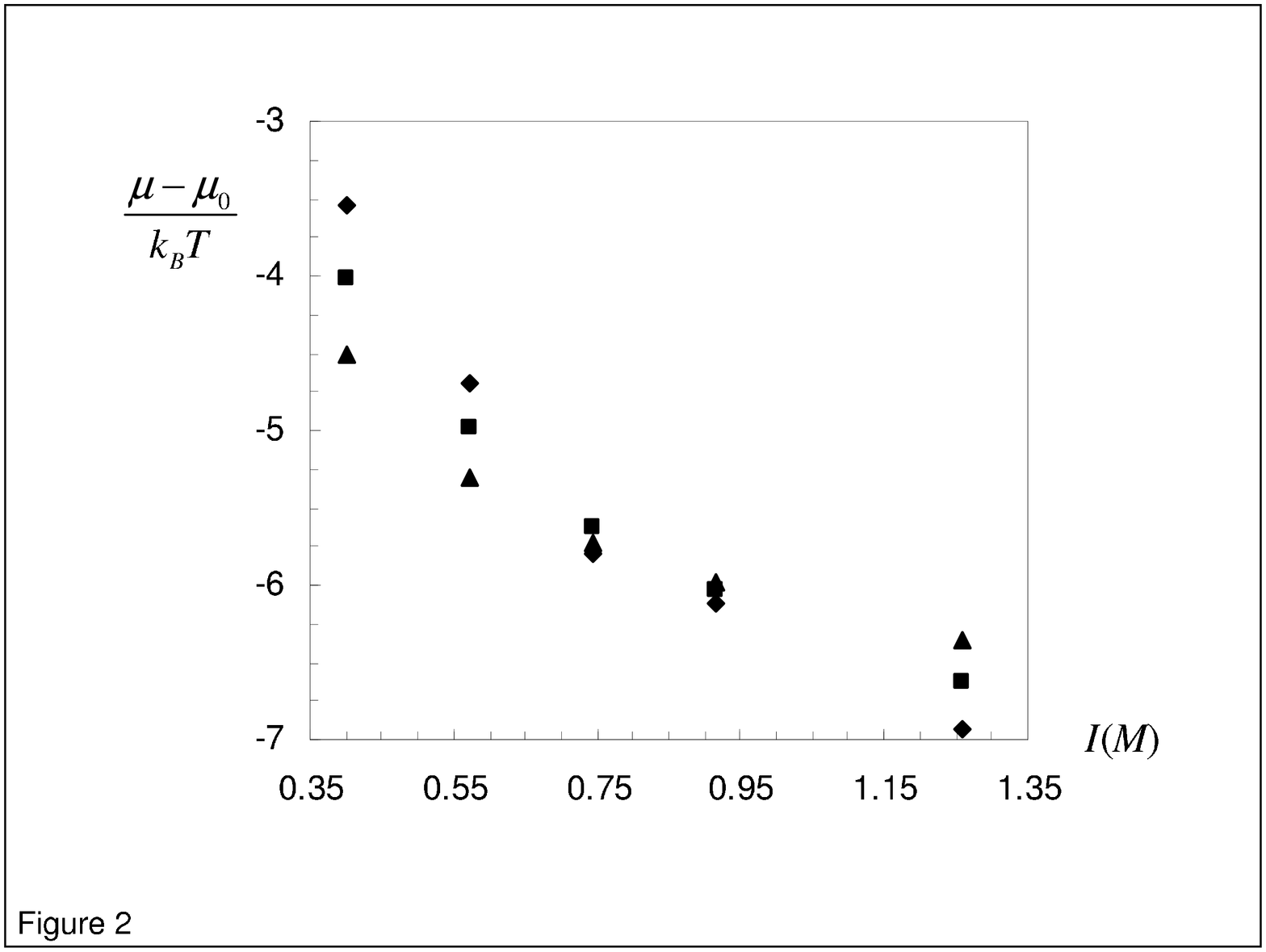, scale=0.5, angle=270, width=250pt}
    \end{center}
  \end{minipage}
  \begin{minipage}[t]{.45\textwidth}
    \begin{center}
      \epsfig{file=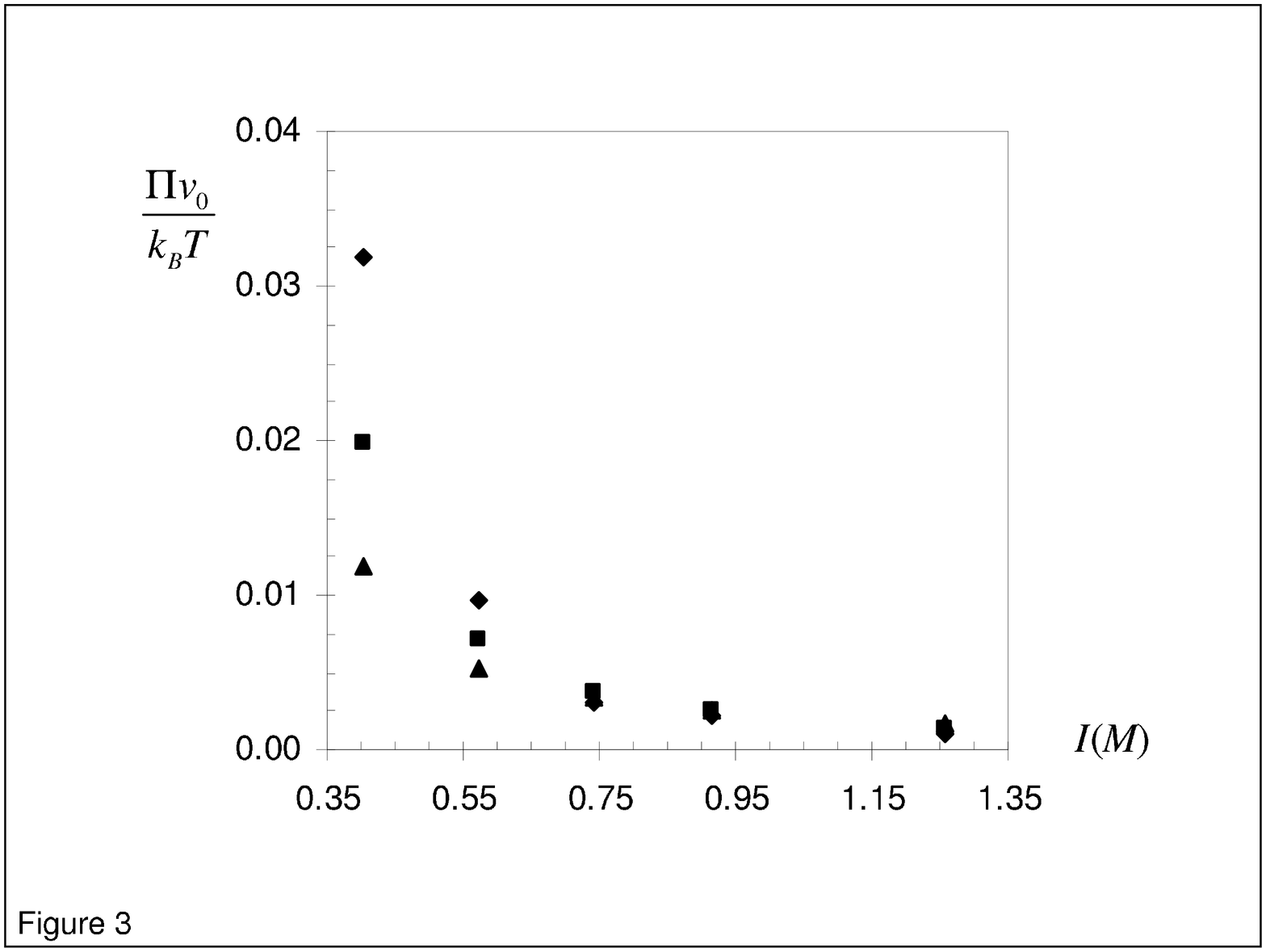, scale=0.5, angle=270, width=250pt}
    \end{center}
  \end{minipage}
  \begin{minipage}[t]{.45\textwidth}
    \begin{center}
      \epsfig{file=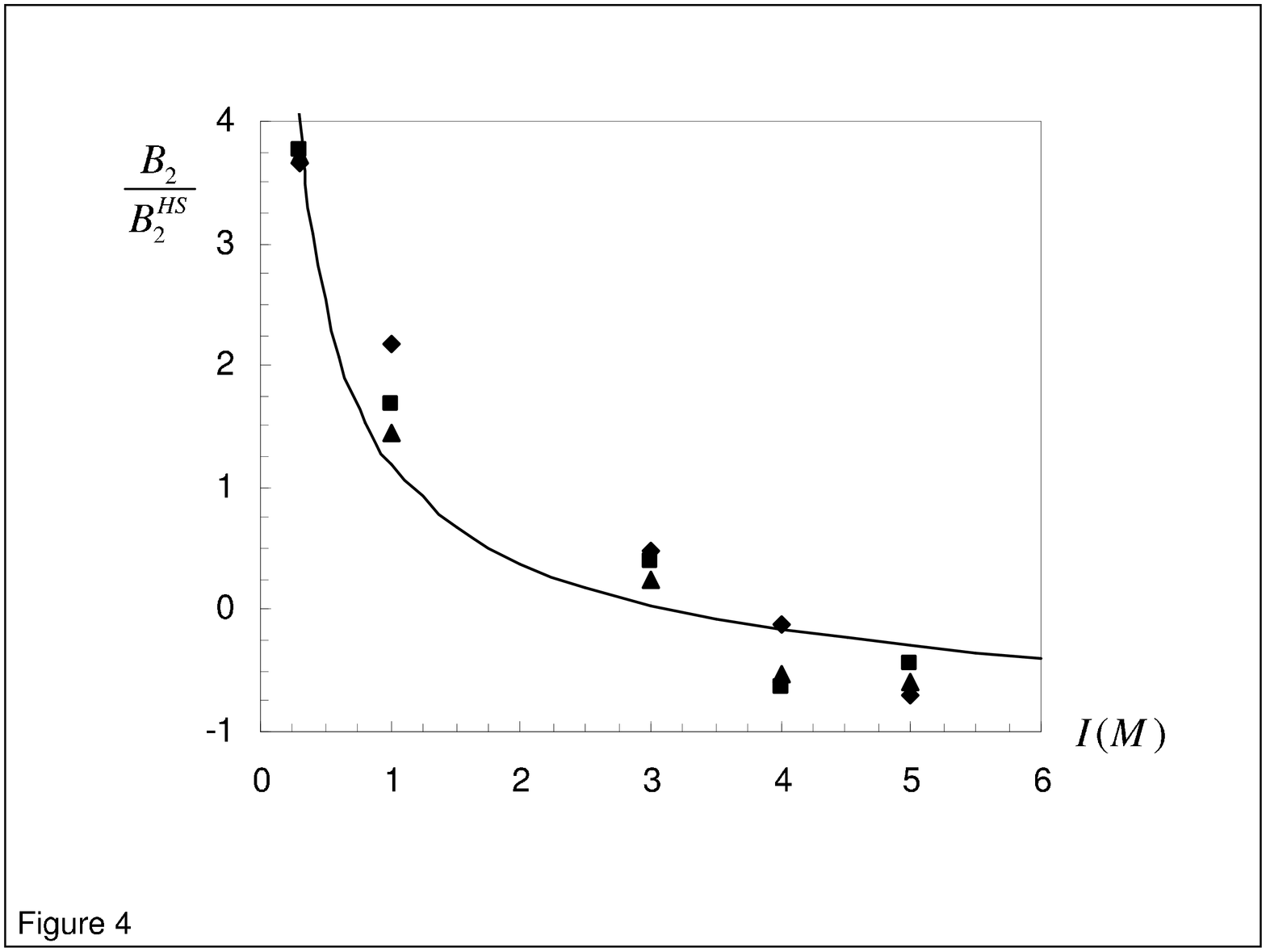, scale=0.5, angle=270, width=250pt}
    \end{center}
  \end{minipage}
  \begin{minipage}[t]{.45\textwidth}
    \begin{center}
      \epsfig{file=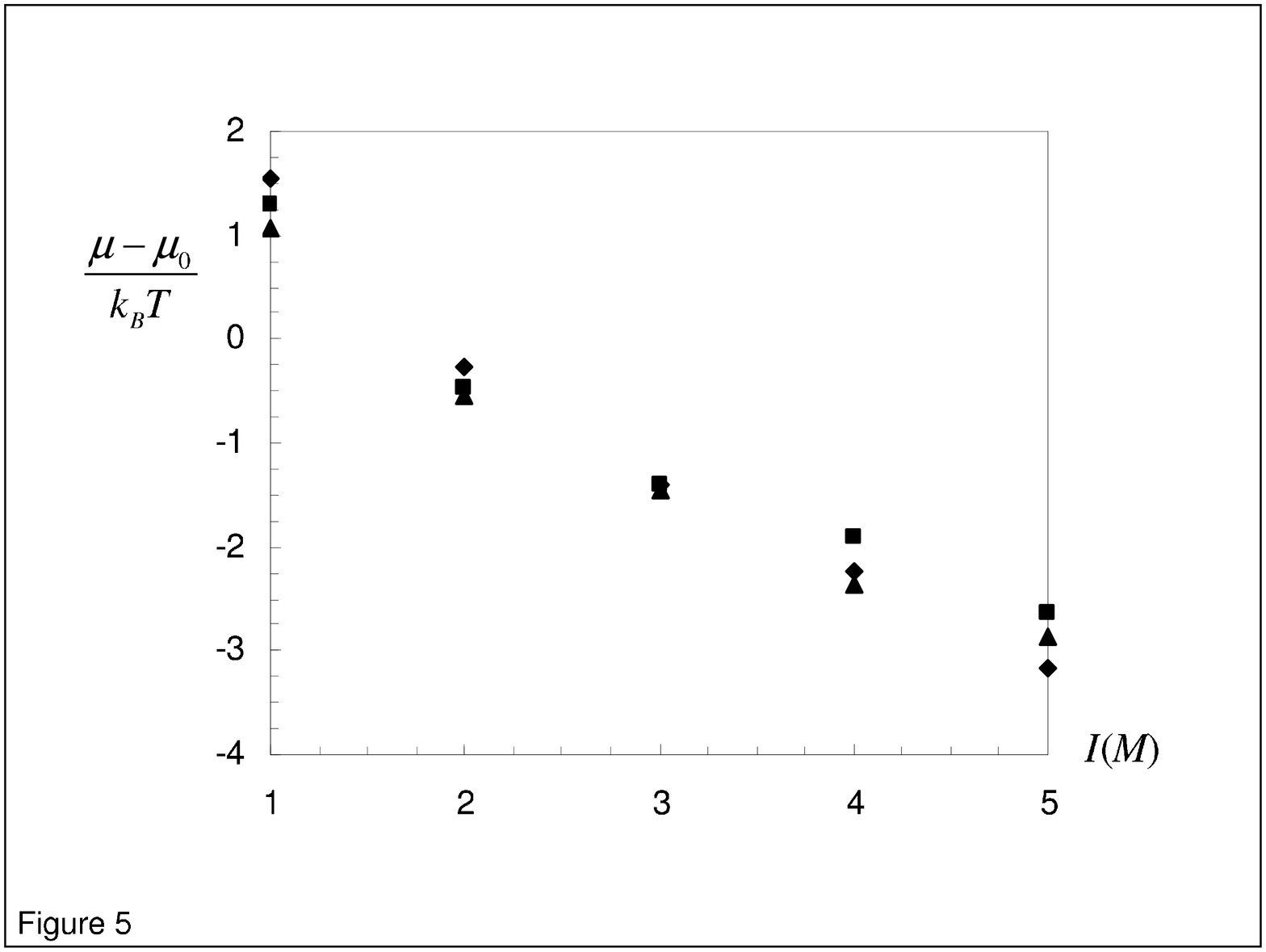, scale=0.5, angle=270, width=250pt}
    \end{center}
  \end{minipage}
  \begin{minipage}[t]{.45\textwidth}
    \begin{center}
      \epsfig{file=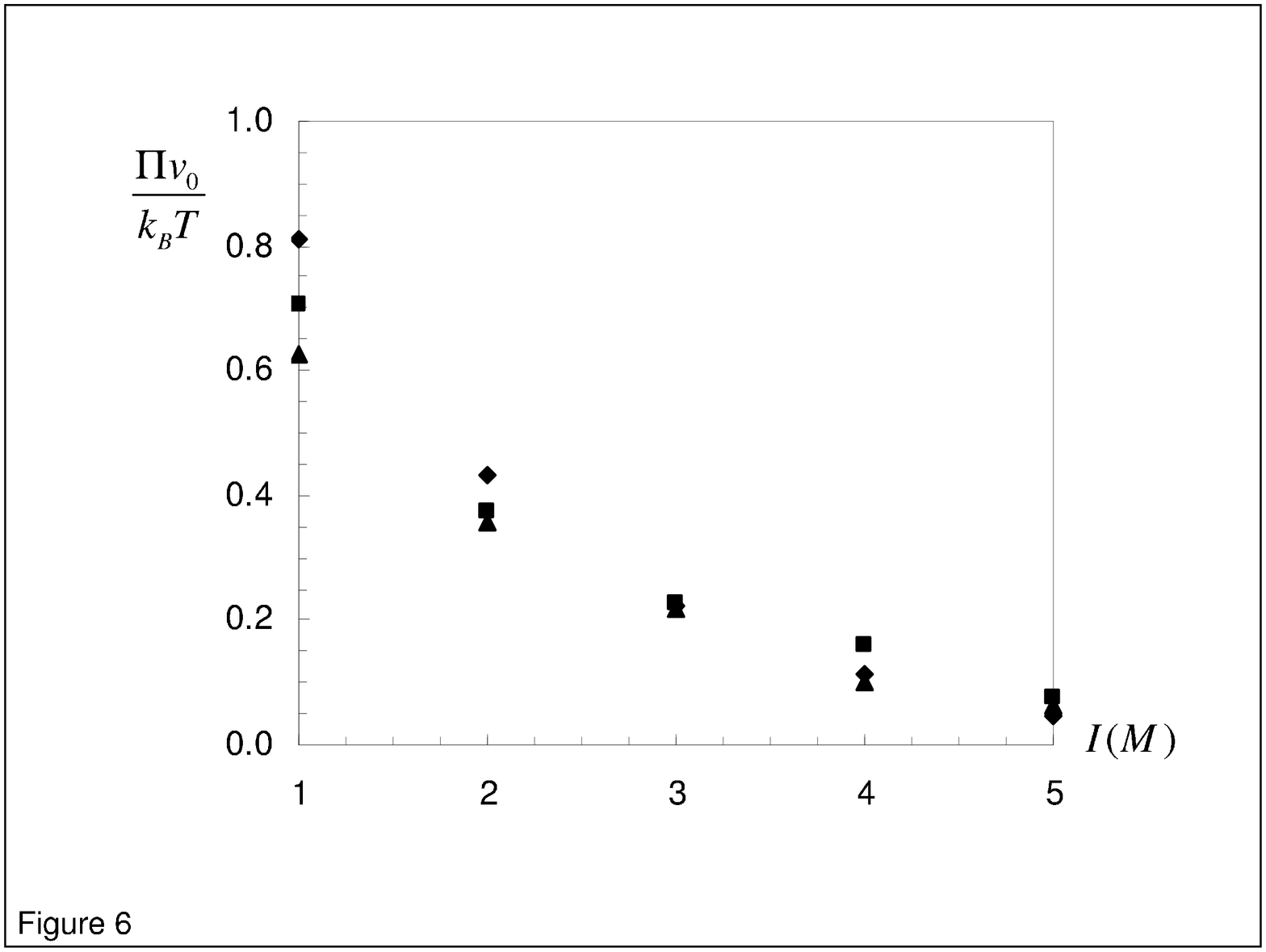, scale=0.5, angle=270, width=250pt}
    \end{center}
  \end{minipage}
  \end{figure}
  \begin{figure}[h]
  \begin{minipage}[t]{.45\textwidth}
    \begin{center}
      \epsfig{file=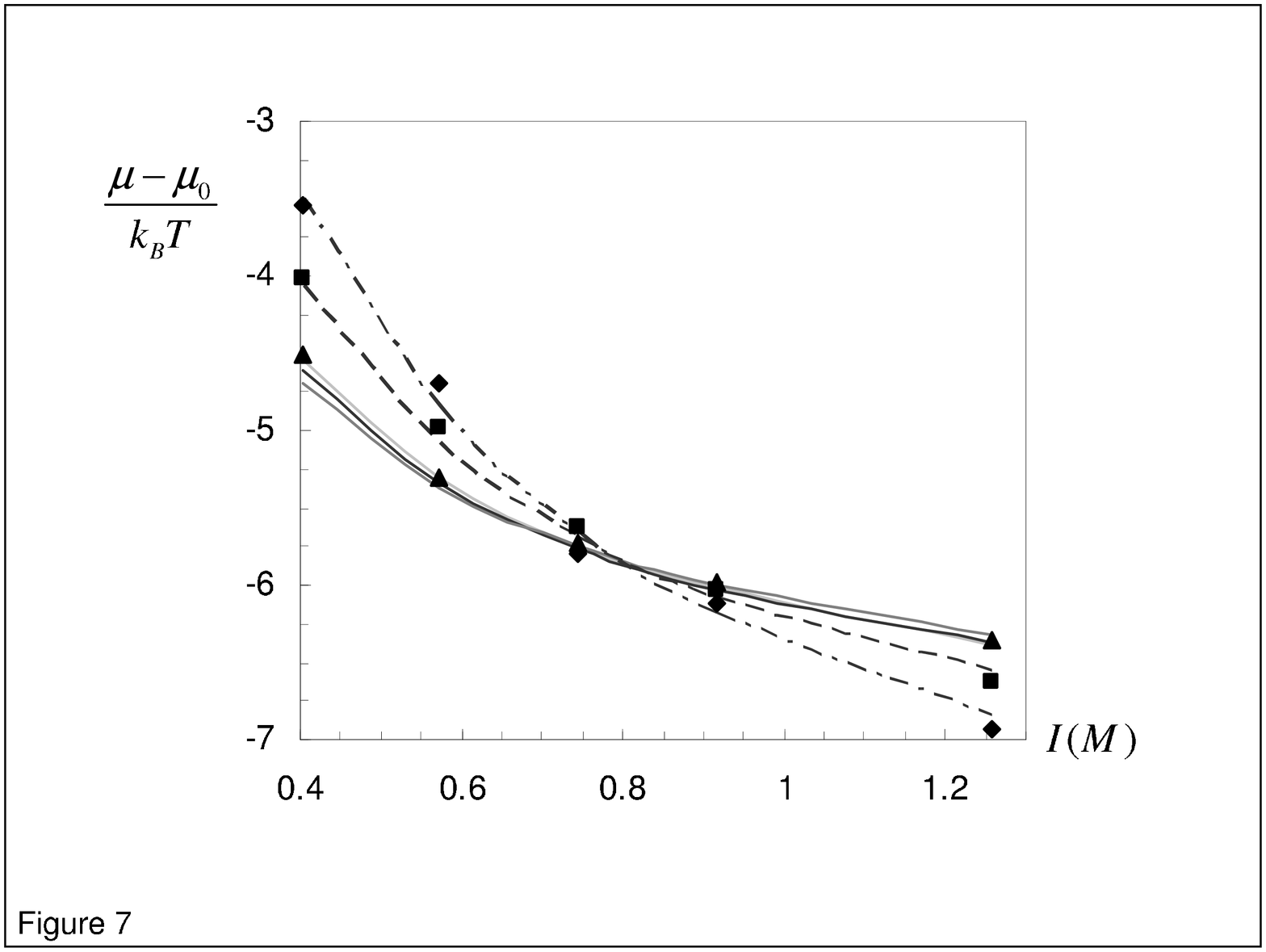, scale=0.5, angle=270, width=250pt}
    \end{center}
  \end{minipage}
  \begin{minipage}[t]{.45\textwidth}
    \begin{center}
      \epsfig{file=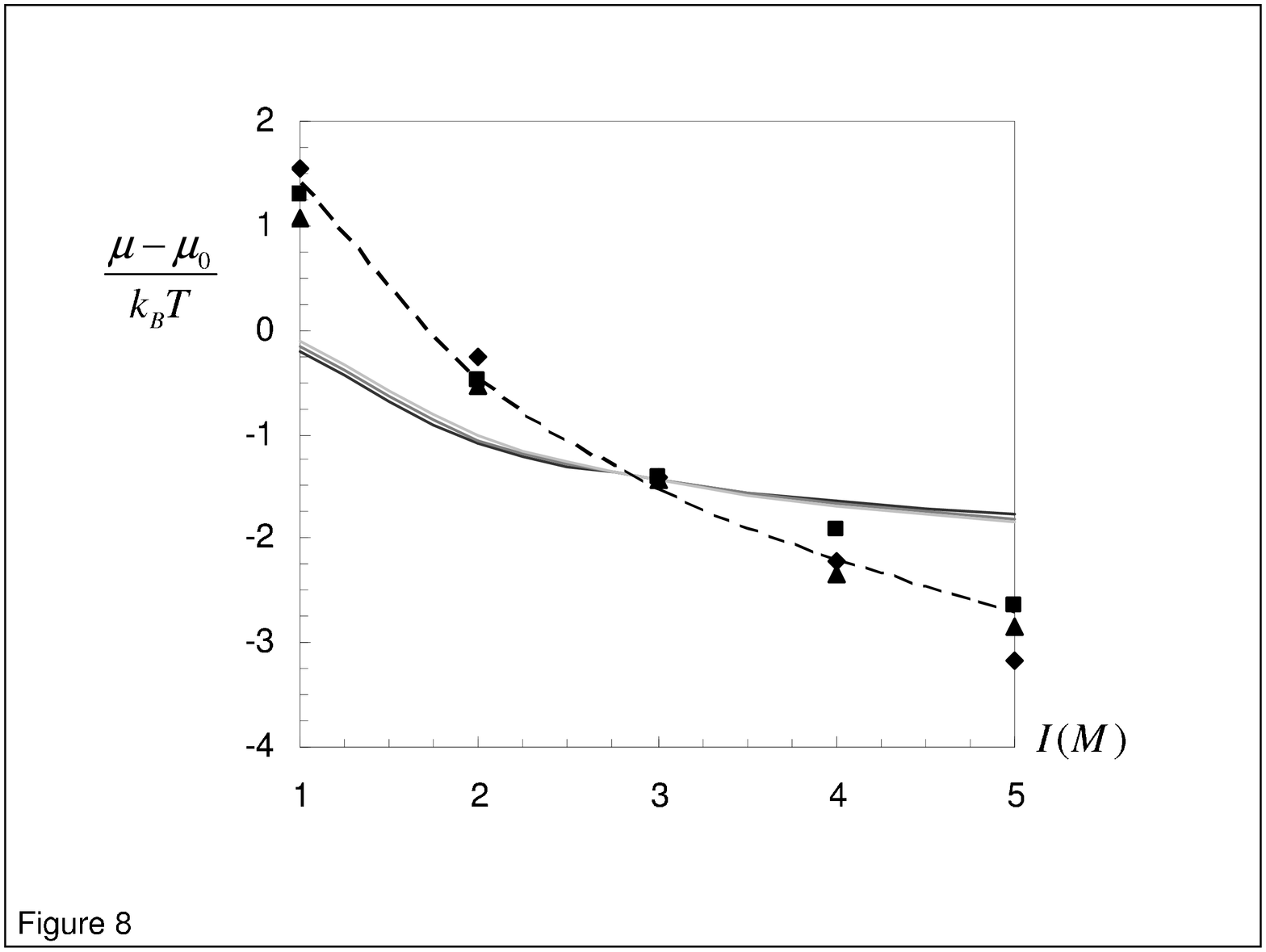, scale=0.5, angle=270, width=250pt}
    \end{center}
  \end{minipage}
  \begin{minipage}[t]{.45\textwidth}
    \begin{center}
      \epsfig{file=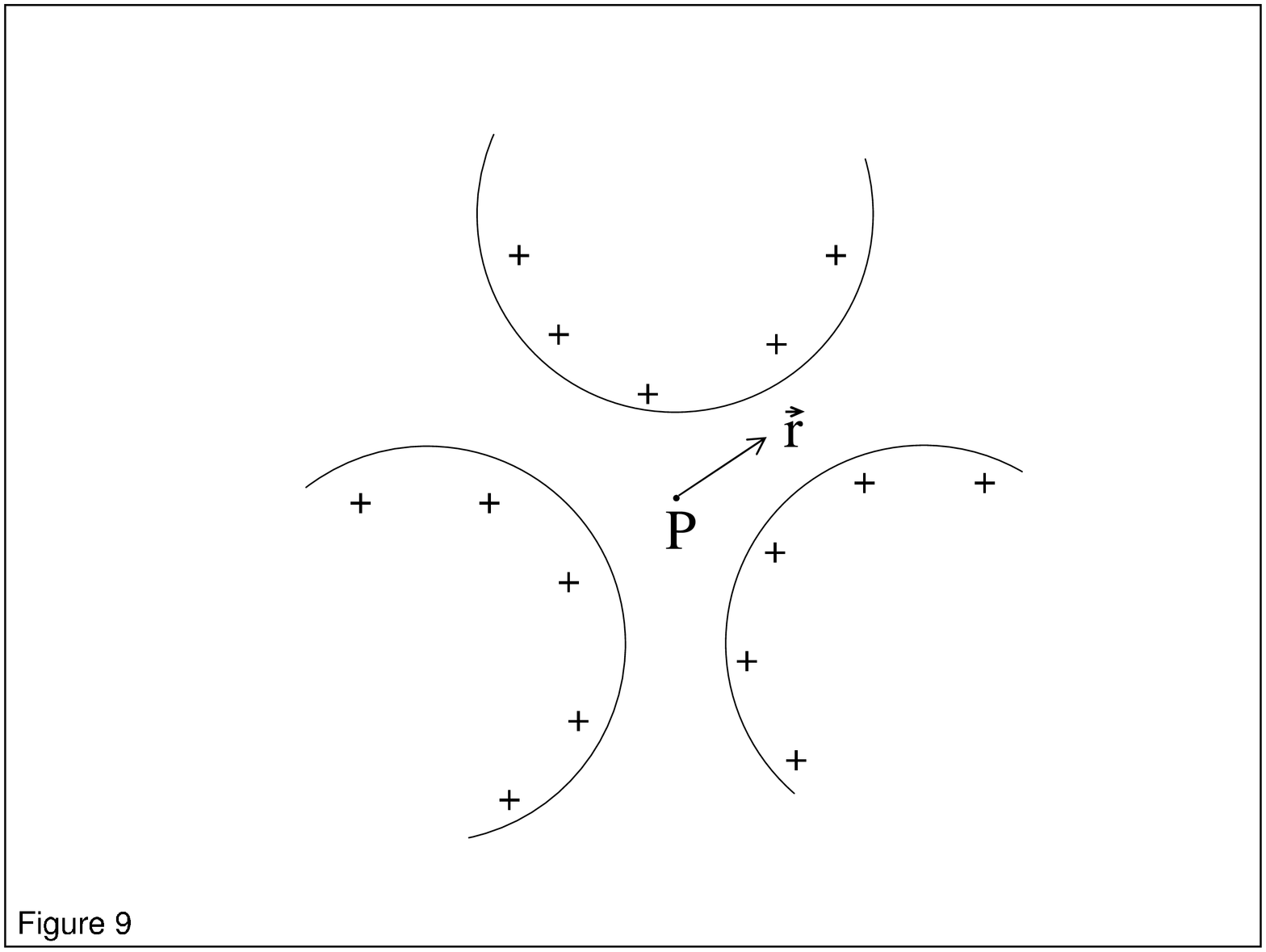, scale=0.5, angle=270, width=240pt}
    \end{center}
  \end{minipage}
  \end{figure}

\end{document}